\documentclass[pre,twocolumn,superscriptaddress,amsmath,amssymb,aps]{revtex4-2}

\usepackage{graphicx,color,comment,xcolor}
\usepackage{dcolumn}
\usepackage{bm}
\usepackage{bibunits}
\defaultbibliography{feedback}
\defaultbibliographystyle{apsrev4-2}
\usepackage{lineno}

\begin{document}

\title{Feedback percolation on complex networks}

\author{Hoseung Jang}
\affiliation{Advanced-Basic-Convergence Research Institute, Chungbuk National University, Cheongju, Chungbuk 28644, Korea}
\author{Ginestra Bianconi}
\email{ginestra.bianconi@gmail.com }
\affiliation{Centre for Complex Systems, School of Mathematical Sciences, Queen Mary University of London, Mile End Road, London, E1 4NS, UK}
\author{Byungjoon Min}
\email{bmin@cbnu.ac.kr}
\affiliation{Advanced-Basic-Convergence Research Institute, Chungbuk National University, Cheongju, Chungbuk 28644, Korea}
\affiliation{Department of Physics, Chungbuk National University, Cheongju, Chungbuk 28644, Korea}

\date{\today}

\begin{abstract}
Traditional percolation theory assumes static microscopic rules,  
limiting its ability to describe real-world complex systems where macroscopic 
order actively regulates local interactions. Here, we introduce feedback 
percolation, a minimal framework that dynamically couples the microscopic activation 
probability to the macroscopic size of the giant component. We show
that this simple feedback mechanism produces a rich variety of  behaviors 
both analytically and numerically. 
Depending on the feedback functions, the system exhibits discontinuous 
jumps, hybrid transitions, limit-cycle oscillations, and routes to chaos, absent in classical percolation. 
Our findings establish that the feedback provides a simple and effective physical mechanism 
for phenomena ranging from self-regulating oscillations to systemic collapse.
\end{abstract}

\maketitle

\section{\label{sec:Intro}Introduction}
Percolation plays a fundamental role in the study of complex systems and their underlying network structure. 
In its classical formulation, large scale connectivity, characterized by the size of 
the giant component, is determined by a fixed occupation probability $p$ of sites or bonds.
This model has been successful in describing a wide range of phenomena, from
fluid flow in porous media \cite{broadbent1957,stauffer2018} and the conductivity
of disordered materials \cite{kirkpatrick1973} to the spread of infectious
diseases \cite{cardy1985epidemic,newman2002spread}. 
By characterizing the network connectivity  \cite{stauffer2018}, percolation offers key theoretical insights into system robustness \cite{albert2000error,buldyrev2010catastrophic,bonamassa2025hybrid,artime2024robustness},  quantum communication performance \cite{meng2023percolation,meng2025path,cirigliano2026dynamical}
and the topological requirements for the emergence of ordered states and phase transitions \cite{dorogovtsev2008critical,bianconi2018multilayer,araujo2014recent,lee2018recent,cho2026recent,christensen2005}. 
Traditional percolation models
are ``static'' because the activation probability $p$ is fixed and the activation
is permanent \cite{stauffer2018}.
Once active nodes or links are known, the connected clusters are determined entirely by the topological properties of the network.

In many real-world complex systems, local activation probability and macroscopic size
of the giant component often exhibit a coupled influence, indicating the presence of
dynamic feedback \cite{buldyrev2010catastrophic,son2012percolation,sun2023dynamic,zhao2013percolation,hebert2025self}.
A prominent and extensively studied approach to such coupling is cascading failures in interdependent
networks \cite{buldyrev2010catastrophic,parshani2010interdependent,baxter2012avalanche,gao2012networks},
where the failure of nodes in one network propagates to dependent nodes in another
and a node survives only if it remains part of the giant component.
In this way the macroscopic connectivity itself becomes coupled to the microscopic survival
of nodes, producing an abrupt collapse \cite{buldyrev2010catastrophic,baxter2012avalanche,zhou2014simultaneous}.
A similar mechanism operates within a single network, where mutual dependencies
can trigger a cascade leading to an abrupt collapse \cite{bashan2011percolation,kuehn2021universal}.
A related mechanism arises in triadic percolation \cite{sun2023dynamic,millan2025topology,sun2024higher,sun2026triadic,millan2024triadic,aghaei2026superstable},
where regulatory nodes activate or deactivate links, which leads to a dynamical giant component
whose size can undergo not only discontinuous phase transitions and period-two bifurcations but also a route to chaos.
Suppression of the giant component was also examined in explosive percolation,
where a local selection rule delays the emergence of the giant component \cite{achlioptas2009explosive,dsouza2019explosive}.
Beyond percolation, feedback between a control parameter and the order parameter has been
explored across a range of phase transition models, from competitive multilayer dynamics \cite{danziger2019dynamic}
to interdependent Ising systems \cite{bonamassa2021realizing,gross2024microscopic} and
their experimental realization in disordered superconductors \cite{bonamassa2023interdependent}.
Related feedback mechanisms have been studied in the context of spin systems \cite{ma2025dynamics,ma2025mixed}
and quantum communication \cite{meng2025quantum}.

\begin{figure*}
\centering
\includegraphics[width=\textwidth]{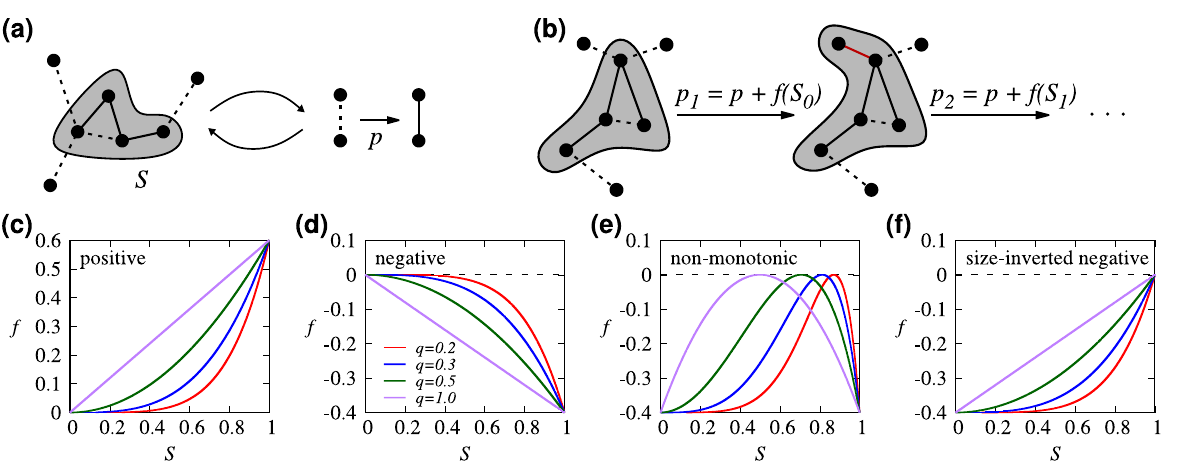}
\caption{\label{fig1}
Feedback mechanisms and functions.
(a) Schematic illustration of feedback in percolation models. 
The activation probability and the size of the giant component influence each other. 
(b) An iterative process of the feedback mechanism where the activation probability is updated as $p_{n}=p+f(S_{n-1})$. 
Here, $p_n$ and $S_n$ represent the activation probability and the size of the giant component after $n$ iterations of the feedback process, respectively. 
Solid and dotted lines indicate active and inactive links, respectively, and the largest component is highlighted in gray. The newly activated link due to the feedback is marked in red.
(c-f) Feedback function $f$ as a function of $S$ for the positive, negative, non-monotonic, and size-inverted negative 
cases, respectively, for the bare probability $p=0.4$. 
Curves with different colors correspond to feedback strength $q=0.2, 0.3, 0.5$, and $1.0$.
}
\end{figure*}

In this work, we introduce a simple and minimal percolation-based framework in which the feedback between
the macroscopic order and the microscopic activation of links is given by a single scalar function
of the giant component size.
Previous models often encode feedback through specific microscopic dependency \cite{buldyrev2010catastrophic,bashan2011percolation} or regulatory rules \cite{sun2023dynamic}. In the present model, this structure is replaced by an explicit scalar feedback function that captures the general mechanisms characterizing feedback by directly coupling the occupation probability to the giant-component size.
For example, in epidemic spreading, where the prevalence of a disease influences individual behaviors,
such as improved personal hygiene and social distancing, thereby reducing the local
transmission probability of the pathogen \cite{funk2009spread}.
Another example can be found in neural systems in which Hebbian plasticity dictates that the strength
and probability of connections are reinforced by mechanisms that ultimately depend on the collective activity of
the network \cite{hebb2005,rapisardi2022}.
Quantum communication networks enhanced by distributed quantum memories are also an example of dynamic
feedback \cite{meng2025quantum}.
Such a modeling is also valuable when there are not enough data
about the actual structure of microscopic mechanisms and can therefore be described directly at the macroscopic level.
In these cases, the macroscopic order is not merely a consequence of the activation probability of nodes or links; it actively shapes the activation probability, forming a feedback loop.

Within this framework, the activation probability is updated dynamically according to the size of the giant
component, allowing the network to continuously adapt its local connectivity in response to its macroscopic state.
This model aims to capture how such
a feedback mechanism influences the phase transitions of percolation on complex networks, and
to provide an effective theoretical framework to understand dynamical percolation.
We establish self-consistency equations to analyze the iterative evolution of
the system, and we find that the feedback qualitatively alters the nature of
percolation transitions. Specifically, it gives rise to a rich variety of
complex behaviors, including discontinuous jumps, stable oscillatory states,
double transitions, and route to chaos, showing that feedback is a fundamental driver of diverse
percolation transitions.
This work broadens the scope of generalized percolation models by capturing diverse percolation
transitions through diverse forms of feedback functions.

\section{\label{sec:Results}Results}
\subsection{\label{sec:Model}Feedback percolation}

We consider a network of size $N$, which is the number of nodes, where all links are initially inactive. 
In the initial step, say $n=0$, each link is activated with a bare probability $p_0=p$. 
We identify the connected components and measure the size of the giant component, $S_0$. 
These procedures are identical to ordinary bond percolation on networks. 
We then incorporate the feedback effect between the control and order parameters, $p$ and $S$, as shown in Fig.~\ref{fig1}(a). 
The feedback function is encoded in the activation probability, which depends on the size of the giant component. 
Specifically, the feedback mechanism determines the bond occupation probability $p_1$ at step $1$ depending on the size $S_0$ of the giant component at step $n=0$, according to  $p_1 = p + f_1(S_0)$ where $f_1(x)$ is the feedback function at step $1$. 
Subsequently, the links are activated or deactivated to satisfy this new probability $p_1$, and the size of the corresponding giant component will change accordingly. This leads to an iterative update of the bond occupation probability and the corresponding size of the giant component. Specifically, at each step $n$, the activation probability $p_n$ is given by 
\begin{align}
\label{Eq_pn}
p_{n}=p + f_{n}(S_{n-1}),
\end{align} 
where $f_n(x)$ is the feedback function at step $n$, while  the size of the giant component $S_n$ is determined by the bond occupation probability $p_{n}$ as in standard bond percolation [see Fig.~\ref{fig1}(b)].

Depending on the choice of the feedback function, $f_n(S)$ one can observe different critical and dynamical behavior of the giant component. For the sake of simplicity, in this work we consider feedback functions that do not depend on the iteration step $n$, implying that $f_n(S)=f(S)$ for every value of $n$. 
In this case, $S_n$ is determined by $p_n$ that on its turn is determined by $S_{n-1}$. Thus,  these iterative equations can be cast into a one-dimensional map \cite{strogatz2024nonlinear}, $S_n=h(S_{n-1})$, with $h(x)$ being independent of $n$. Therefore, in feedback percolation the giant component can become dynamical, potentially leading not only to discontinuous phase transitions but also to period-two oscillations and even a route to chaos, similarly to what happens for triadic percolation \cite{sun2023dynamic}. 
As we will see in the following, the specific choice of the feedback mechanism adopted in this work, leads either to a fixed steady state, to a period-two oscillation away from its transient behavior, or to a route to chaos.
In the case in which the iteration  reaches a steady state, meaning that 
 $S_n=S_{n-1}$, we  measure the final fraction of nodes in the giant component $S_\infty$ 
with respect to the bare probability $p$. In the case in which the iteration reaches a stable period-two oscillation, i.e., $S_{n}=S_{n-2}$, we will measure the fraction of nodes in the giant component $S_\infty$ on both even and odd iterative steps.
In more complex dynamical situations, where the dynamics has a longer oscillatory period or becomes chaotic, we measure the entire timeseries $S_n$ of the size of the giant component at step $n$ after a transient period.

According to the mathematical form of $f(S)$, the feedback function can cause positive or negative feedback effect. 
Here, the terms positive and negative refer to whether the large-scale giant component reinforces or suppresses further link activation. 
We implement the effect of feedback functions depending on the size of the giant component adopting power-law functional expressions of the type $S_n^{1/q}$.
Here, the parameter $q$ determines the strength of the feedback effect. 
on the bond percolation with a feedback mechanism meaning 
that links are dynamically activated or deactivated depending on the size of the giant component. 
Specifically, we consider four forms of the feedback function $f(S)$, illustrated in Figs.~1(c-f): 
the positive feedback $f(S)=(1-p)S^{1/q}$, which reinforces link activation as the giant component grows; 
the negative feedback $f(S)=-pS^{1/q}$, which suppresses it; 
the non-monotonic feedback $f(S)=-p(2S^{1/q}-1)^{2}$, which combines reinforcement 
and suppression depending on the giant component size; 
and the size-inverted negative feedback $f(S)=-p(1-S^{1/q})$, which suppresses activation 
most strongly when the giant component is small.

We study the feedback percolation model on Erd\H{o}s-R\'enyi (ER) graphs with average degree $z$. 
Throughout, we use $z=4$ and the network size $N=10^5$. 
Note that for such a graph, a finite fraction of nodes remains outside the giant component, 
so that even at $p=1$ the size of the giant component is slightly below unity, with $S \approx 0.98$ at $p=1$.
In the following subsections, we analyze the feedback percolation model for the four feedback functions: 
positive, negative, non-monotonic, and size-inverted negative feedback.

\begin{figure*}[t]
\centering
\includegraphics[width=\textwidth]{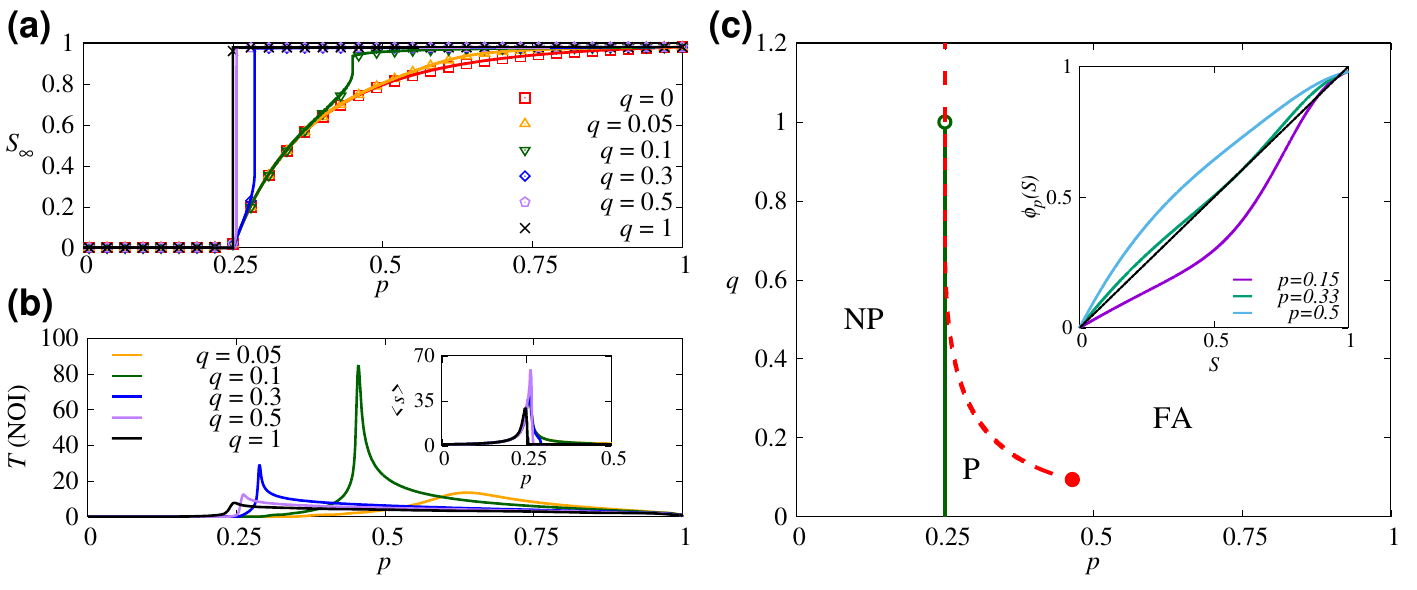}
\caption{\label{fig_Positive}
Diverse transitions under positive feedback.
(a) The size of the giant component, $S_\infty$, under positive feedback as a function of probability $p$ on Erd\H{o}s-R\'enyi (ER) graphs with average degree $z=4$. Symbols represent numerical results for network size $N=10^5$, and solid lines represent theoretical predictions. 
(b) The number $T$ of iterations required to reach the steady state in numerical simulations as a function of $p$. The inset displays the average size of finite connected components. 
(c) Phase diagram in the $(p,q)$ plane for the positive feedback model. The solid line indicates the location of continuous percolation transitions, while the dashed line indicates that of a discontinuous jump of $S_\infty$. 
Here, $\mathbf{NP}$, $\mathbf{P}$, and $\mathbf{FA}$ denote the non-percolating, percolating, and fully activated phases, respectively.
The inset shows the interaction between $\phi_p(S)$ and size $S$ for feedback strength $q=0.2$. 
}
\end{figure*}

\subsection{Diverse percolation transitions in positive feedback}

In this section, we analyze the percolation model under positive feedback, $f(S_n)=(1-p) S_n^{1/q}$, by varying the initial activation probability $p$ and feedback strength $q$. Due to the positive feedback, the activation probability $p_n$  increases, leading to the activation of links that were inactive in the previous step.
Figure~\ref{fig_Positive}(a) shows the size $S_\infty$ of the giant component as a function of $p$ for various values of feedback strength $q$ on ER graphs with $z=4$. We confirm that the theoretical predictions (lines) agree well with the numerical results (symbols). We can identify the percolation threshold as $p_c=1/z$ where the giant component appears. When the size of the giant component at the initial stage $S_0$ is zero, there is no effect of the positive feedback. Therefore, the location of the percolation threshold under the positive feedback remains the same as in ordinary percolation.

The effect of the positive feedback, however, plays a role when $S_0$ is non-zero. We found that positive feedback can generate a ``discontinuous'' increase in the size $S_\infty$. The location $p_d$ where the jump occurs can be identified by the condition Eq.~(\ref{Eq_pd}). We found that the location $p_d$ gets closer to $p_c$ as the strength $q$ increases. 
When $q=1$, the location of the jump $p_d$, and percolation threshold $p_c$ coincide, meaning that the emergence of $S_\infty$ becomes discontinuous at $p_c$. 
The discontinuous transition at $p_d$ exhibits a hybrid nature, which is followed by a scaling behavior, $(S_d - S_\infty) \sim (p_d - p)^{1/2}$, for $p < p_d$, where $S_d$ denotes the size of the giant component at the jump point $p_d$ \cite{zhou2014simultaneous,kuehn2021universal} (see  Supplementary Information for the details).
The increase of $p_n$ is accelerated with an increase in the size of the giant component; once the jump occurs, the system reaches {a highly activated state.

The two transitions at $p_c$ and $p_d$ are driven by distinct mechanisms. At the continuous transition, the giant component emerges via ordinary percolation independent of the iterative feedback process. Therefore, the percolation transition where the giant component first appears is induced by the critical behavior of finite component size distributions, showing a divergence of the average size $\langle s \rangle$ of finite components at $p_c$ [see the inset of Fig.~\ref{fig_Positive}(b)]. In contrast, the discontinuous transition from the percolating to the fully activated phase is driven by repeated updates of $p$ and $S$ due to the feedback. 
This feedback-driven iterative process is highlighted by the number of iterations (NOI), $T$, defined as the number of update steps required for the activation probability to converge. As shown in Fig.~\ref{fig_Positive}(b), the NOI, $T$, exhibits a sharp peak at the point of the discontinuous transition to the fully activated phase that diverges.
These results clearly show that the discontinuous jump arises from the divergence of 
iteration steps, rather than the geometric transition characteristic of ordinary percolation.

The phase diagram of the positive feedback model is depicted in Fig.~\ref{fig_Positive}(c). We identified three distinct phases: non-percolation (NP), percolating (P), and fully activated (FA). 
In the positive feedback, increasing $p$ can lead to the occurrence of double transitions
which refer to the sequential occurrence of a continuous transition at $p_c$ followed by a discontinuous jump at a higher value $p_d$. This discontinuous jump eventually vanishes at the critical point $(p_{cp}, q_{cp}) \approx (0.464, 0.094)$.
In addition, as the two transition lines meet at $(p_{ce}, q_{ce}) = (p_c, 1)$, forming a critical endpoint where the double transitions merge into a single transition, and the percolation transition shifts from continuous to discontinuous.
The inset of Fig.~\ref{fig_Positive}(c) shows the intersections between $\phi_p(S_\infty) = 1 - e^{ - z S_{\infty} \left[ p+(1-p)S_{\infty}^{1/q} \right] }$ and $S_\infty$, corresponding to the steady-state solution. Specifically, in the double transition regime, the curve $\phi_p(S_\infty)$ first continuously intersects the diagonal near the origin at $p_c$, and subsequently undergoes a bifurcation at a larger value of $S_\infty$, 
which triggers the secondary discontinuous jump at $p_d$ as $p$ further increases.

\begin{figure*}
\centering
\includegraphics[width=\linewidth]{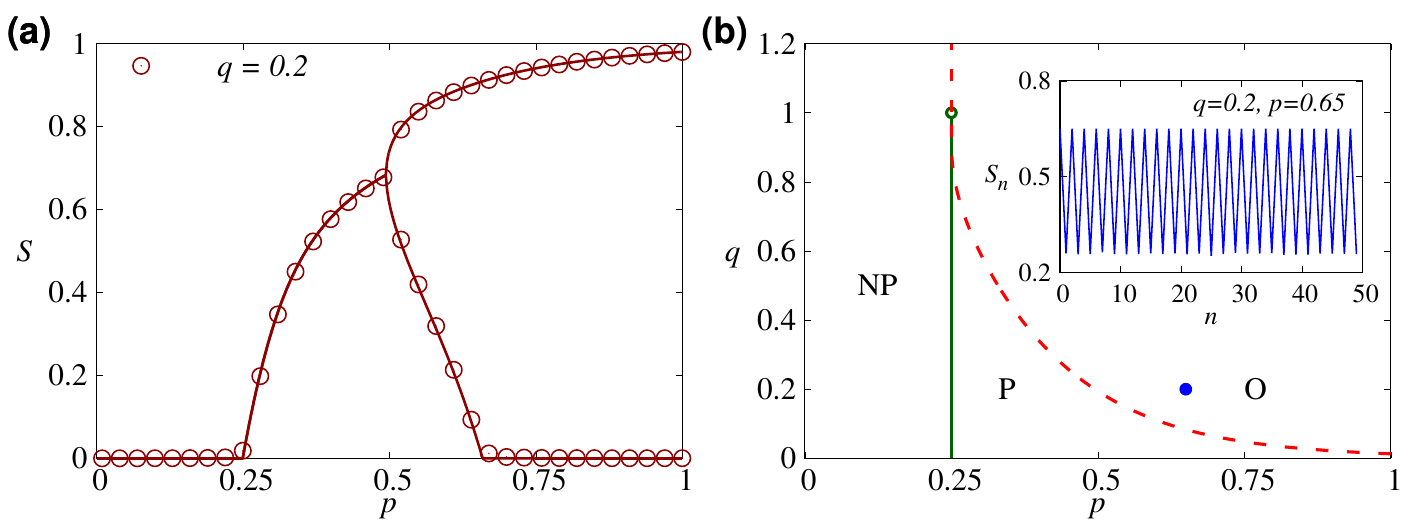}
\caption{\label{fig_Negative}
Emergence of stable oscillations under negative feedback.
(a) Bifurcation diagram under negative feedback with feedback strength $q=0.2$ on Erd\H{o}s-R\'enyi (ER) graphs with average degree $z=4$. Numerical results (symbols) for network size $N=10^5$ and theoretical predictions (lines) are shown together. 
(b) Phase diagram in the $(p,q)$ plane for the negative feedback model. 
The solid green line represents the boundary $p_c$ of the percolation transitions, and the red dashed line indicates the boundary $p_o$ of the onset of oscillatory behavior.
The blue dot marks the $(p,q)$ point at which the inset is plotted.
Here, $\mathbf{NP}$, $\mathbf{P}$, and $\mathbf{O}$ denote the non-percolating, percolating, and oscillatory phases, respectively.
The inset shows the time course of the size $S_n$ of the giant component.
}
\end{figure*}

\subsection{Stable oscillations in negative feedback}

We now turn to the negative feedback model and examine how negative feedback generates stable oscillations on ER graphs. 
In the negative model, the term $f(S_{n-1})= -p S_{n-1}^{1/q}$ governs the strength of the feedback effect. As $S_{n-1}$ increases, the negative feedback actively suppresses the formation of the giant component, effectively reducing its size. The exponent $q$ controls the sensitivity of this suppression: as $q$ increases, the negative feedback becomes more pronounced even for small values of $S_{n-1}$.

We can predict the location of the percolation threshold as $p_c = 1/z$ as in ordinary bond percolation, and a non-zero fixed point $S^*$ emerges for $p > p_c$.
The stability of the non-zero fixed point $S^*$ is determined by the linear stability analysis, expressed by Eq.~(\ref{Eq_po}).
The condition leads to the location where the fixed point loses its stability:
\begin{align}
\frac{z p_o (1-S^*){S^*}^{1/q}}{q \left[1-z p_o(1-{S^*}^{1/q})(1-{S^*}) \right] } =1,
\end{align}
where $p_o$ denotes the critical boundary where the fixed point $S^*$ loses its stability for a given $q$. 
The instability of $S^*$ for $p>p_o$ leads to the emergence of oscillatory behavior as shown in the bifurcation diagram of Fig.~\ref{fig_Negative}(a) and the time course of $S_n$ in the inset of Fig.~\ref{fig_Negative}(b).
In this regime where $p>p_o$, the negative feedback term $-p S_{n-1}^{1/q}$ becomes 
so dominant that a large giant component at step $n$ drastically suppresses 
its own formation. This weakens the feedback for the following step, 
allowing the giant component to grow again. Consequently, the system enters a limit 
cycle characterized by the oscillation of  the size of the giant component  between two values, a 
phenomenon driven by the delayed negative feedback mechanism.

The phase diagram is characterized in the parameter space of the occupation probability $p$ and the feedback sensitivity $q$, as illustrated in Fig.~\ref{fig_Negative}(b). We can identify three distinct phases: non-percolating (NP), percolating (P), and oscillatory (O) phases. 
The percolation threshold, $p_c$, divides the non-percolating (NP) and percolating (P) phases. Subsequently, the oscillatory boundary, $p_o$, separates the stable percolating phase 
from the oscillatory (O) phase. 
In the intermediate regime ($p_c < p < p_o$), the system converges to a single non-zero 
value of $S^*$. 
In this regime, the negative feedback is insufficient to destabilize the giant component, and the network maintains a steady-state giant component.
Beyond this bifurcation point, 
the system enters the oscillatory region ($p > p_o$) and results in period-two 
oscillations due to the feedback effect. 
The two lines meet at $(p,q) = (p_c,1)$, and when $q \ge 1$ the percolating phase vanishes and only the oscillatory phase exists for $p\ge p_c$.

The oscillatory behavior identified in this model corresponds to  
self-regulating systems where the growth of a macroscopic structure 
triggers its own suppression. An example can be found in the dynamics of 
epidemic spreading coupled with social policy. When the infection cluster
reaches a critical threshold, it often brings out the implementation of 
negative feedback in the form of social distancing or lockdown measures. 
These interventions effectively fragment the contact network, causing $S$ to shrink. 
As the threat diminishes, policies are relaxed, leading to a new wave of growth. 
In communication networks, congestion control protocols act as a regulatory feedback. 
As data flows aggregate into a large stream, the resulting congestion 
forces a reduction in transmission rates, creating an oscillation 
between high and low utilization. 
Oscillatory phenomena in complex systems such as periodic bursting of financial bubbles, cyclical 
forest fires in ecosystems, and the regulation of synchronization in neural networks
may share a similar underlying mechanism.

\begin{figure}
\centering
\includegraphics[width=8.0cm]{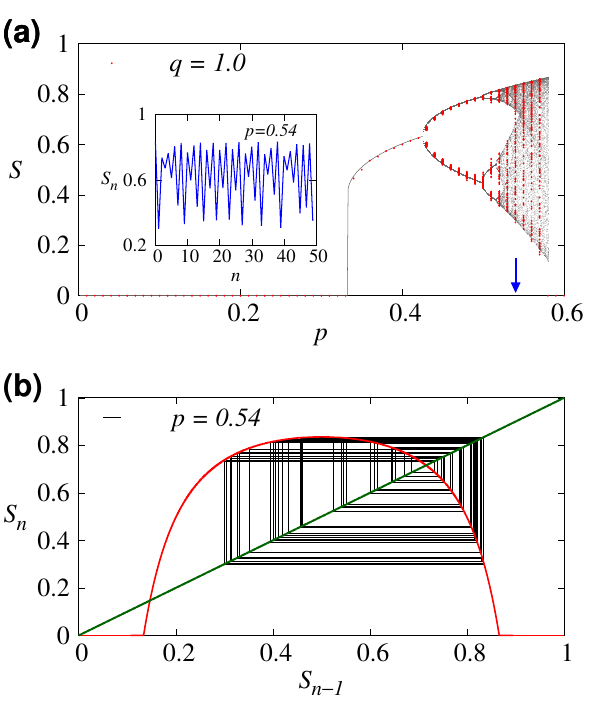} 
\caption{\label{fig_Chaos}
Route to chaos under non-monotonic feedback.
(a) Bifurcation diagram under the non-monotonic feedback with feedback strength $q=1$ on Erd\H{o}s-R\'enyi (ER) graphs with average degree $z=4$. Numerical results (red) for network size $N=10^5$ and the theoretical results (gray) are shown together. The blue arrow marks the value of probability $p$ at which the inset is plotted. The inset shows the time course of the size $S_n$ of the giant component.
(b) Cobweb plot within a chaotic regime: $(p,q)=(0.54,1)$. The red line represents the iterative map, and the green line represents $S_n=S_{n-1}$.
}
\end{figure}

\subsection{Route to chaos in non-monotonic feedback}

We next consider a non-monotonic feedback function, specifically $f(S_{n-1})=-p (2S_{n-1}^{1/q}-1)^2$, and thus the occupation probability at step $n$ becomes $p_n=4 p S_{n-1}^{1/q}(1-S_{n-1}^{1/q})$.
The size of the giant component $S_n$ at step $n$ is described by the iterative map 
$S_n = 1 - \exp[- 4 z p S_n S_{n-1}^{1/q}(1-S_{n-1}^{1/q})]$. 
The onset of the giant component can be identified by the tangency 
condition: $\frac{d S_n}{d S_{n-1}} = 1$ from Eq.~(\ref{Eq_stability}). Subsequently, as the control parameter $p$ increases further, the stable fixed point loses its stability and bifurcates into a period-two oscillatory state as the negative feedback case. The critical point $p_{o}$  
for the onset of this oscillation is exactly determined 
by the condition: $\frac{d S_n}{d S_{n-1}} = -1$.
An example of the emergence of the giant component and period-two 
oscillation with $q=1$ is depicted in Fig.~\ref{fig_Chaos}(a).

As $p$ is raised beyond $p_o$, the dynamics shows a period-doubling
cascade that leads to chaotic behavior. This chaotic behavior is shown in
the bifurcation diagram [Fig.~\ref{fig_Chaos}(a)] and cobweb plot within a chaotic regime [Fig.~\ref{fig_Chaos}(b)].
The inset of Fig.~\ref{fig_Chaos}(a) shows aperiodic time series of $S_n$.
To quantify and precisely identify the onset of chaotic behavior, we numerically compute
the Lyapunov exponent $\lambda$, which measures the time average of 
logarithmic divergence as: 
$\lambda =\frac{1}{\ell} \sum_{n=1}^{\ell} \ln \left| \frac{\partial S_n}{\partial S_{n-1}} \right|$~\cite{strogatz2024nonlinear},
where $\ell$ is the number of iterations used to the long-time average.
The chaotic regime is identified as the parameter region where $\lambda > 0$, 
i.e., $p>0.521$ with $q=1$ in our example in Fig.~\ref{fig_Chaos}(a).
Our analytical analysis  also reveals that this route to chaos belongs to the logistic map universality class \cite{feigenbaum1978quantitative,strogatz2024nonlinear}
(see Supplementary Information for details).
As $p$ increases, the chaotic attractor expands until the orbit reaches 
a value of $S_{n-1}$ for which the updated $p_n$ drops below the percolation threshold, leading to $S_n=0$ in the thermodynamic limit. 
The dynamics here possesses an absorbing state at $S_n=0$.
Once the orbit reaches this state, the giant component cannot re-emerge and the dynamics stops evolving.
Our finding shows how the non-monotonic feedback brings out aperiodic orbit
and leads to chaotic behaviors.

The non-monotonic nature of the feedback function $f(S_{n-1})$ introduces a competition between growth and suppression of the giant component. Unlike monotonic feedback, where the system typically settles into a stable steady state or a stable oscillation, the quadratic-like form of $p_n$ can lead to aperiodic reconfigurations of the network topology in successive steps due to the system's memory of its previous state $S_{n-1}$. This route to chaos implies that in real-world networked systems governed by non-monotonic constraints, for instance, an over-sized component might trigger a self-inhibiting regulatory response, the long-term predictability of the system's connectivity can be fundamentally limited.

\subsection{Relationship to other dynamical percolation models}

\begin{figure}
\centering
\includegraphics[width=8.0cm]{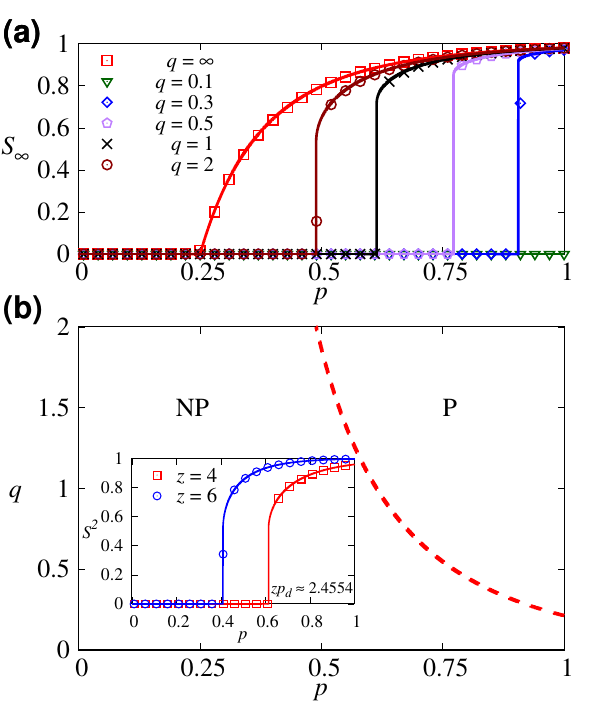}
\caption{\label{fig_Size_inverted_Negative}
Discontinuous transitions under size-inverted negative feedback.
(a) The size of the giant component $S_\infty$ under the size-inverted negative feedback 
as a function of probability $p$ on Erd\H{o}s-R\'enyi (ER) graphs with average degree $z=4$. Theoretical results (lines) show agreement with numerical results (symbols) for network size $N=10^5$. 
(b) Phase diagram in the $(p,q)$ plane for the size-inverted negative feedback model.
Here, $\mathbf{NP}$ and $\mathbf{P}$ denote the non-percolating and percolating phases, respectively.
(inset) The square of the giant component size under the feedback with feedback strength $q=1$, i.e., $p_n=pS_{n-1}$, on ER graphs with $z=4$ and $6$. 
Numerical results (symbols) based on the feedback model with $N=10^5$ show agreement with theoretical results (lines) for the cascading dynamics on interdependent networks. 
}
\end{figure}

We suggest the relationship between feedback percolation and other dynamical percolation models. 
We consider a size-inverted negative feedback function, specifically $f(S_{n-1})=-p(1-S_{n-1}^{1/q})$, and thus the occupation probability at step $n$ is simply $p_n= p S_{n-1}^{1/q}$ on ER graphs.
In this model, the feedback is strongest when the giant component $S_{n-1}$ is small. As $S_{n-1}$ decreases, the term $S_{n-1}^{1/q}$ reduces $p_n$, which suppresses the growth of the giant component. 
As $S_{n-1}$ increases, this suppressing effect becomes weaker. 
For this reason, we call this mechanism size-inverted negative feedback.
We found that the system exhibits a discontinuous phase transition, characterized by a 
sharp jump in the giant component size as shown in Fig.~\ref{fig_Size_inverted_Negative}(a). 
The phase diagram between the percolating and non-percolating phases is depicted 
in Fig.~\ref{fig_Size_inverted_Negative}(b). 
The transition at $p_d$ also shows a hybrid transition \cite{zhou2014simultaneous,cho2026recent}, meaning that 
both a discontinuous jump and a scaling behavior appear.

We further note that although the dynamics of our model and cascading dynamics on interdependent networks
are different, their mean-field self-consistency equations coincide on random networks.
The self-consistency equation of the size-inverted negative feedback case with $q=1$ is given by
$S = 1 - e^{-z(pS)S}$.
By defining a new variable $M=S^2$, with $S \ge 0$, this equation becomes 
$M = (1 - e^{-zpM})^2$ \cite{buldyrev2010catastrophic,bashan2011percolation,gao2012networks,son2012percolation},
which is identical to the self-consistency equation for the cascading failures on two-layer ER networks.
This correspondence does not imply that the two dynamical processes are equivalent; rather,
it offers a complementary, single-network interpretation of the same self-consistency 
equation, consistent with earlier observations \cite{bashan2011percolation}.
To be specific, these two frameworks share the feature that macroscopic connectivity is coupled to microscopic activation. 
In conventional interdependent networks, however
this coupling arises from explicit dependency links between different network layers,
through which failures propagate back and forth across layers.
In contrast, our model describes a single layer network in which the bond occupation probability is globally regulated by a feedback function of the giant component size,
allowing diverse dynamical behaviors depending on the form of the feedback function.
As shown in the inset of Fig.~\ref{fig_Size_inverted_Negative}(b), the numerical simulations based on our feedback mechanism (symbols) are in great agreement with the theoretical predictions (lines), in which the transition point shows the relation $zp_d \approx 2.4554$.

To place this mechanism in a broader context, we observe that this model can also 
be related to the triadic percolation \cite{sun2023dynamic,sun2024higher,sun2026triadic} with negative regulations, 
where the activity of the regulatory nodes associated with each link suppresses the probability that the link is active in turn. 
It is to be noted, however, that the regulatory function considered in triadic percolation has a different functional behavior, which 
for exclusively negative regulation leads to oscillations of the order parameter~\cite{sun2023dynamic}, period-four oscillations (on multiplex networks)~\cite{sun2026triadic}, and a route to chaos (on hypergraphs)~\cite{sun2024higher}, in addition to discontinuous transitions.
A suppression of the giant component also arises in explosive percolation \cite{achlioptas2009explosive,dsouza2019explosive,cho2013avoiding}, 
though there it is imposed by a local selection rule for adding links rather than by macroscopic feedback.

\section{\label{sec:Conclusions}Conclusions}

In this study, we introduced a feedback mechanism into classical percolation
theory, showing that the macroscopic state of a network can actively alter
its local connections, and vice versa. While traditional percolation models 
rely on static activation probabilities, our framework reveals that coupling 
the microscopic probability $p$ to the macroscopic giant component size $S$ 
can fundamentally shift the nature of percolation transitions.
By analyzing specific forms of feedback functions, we uncovered a rich variety of emergent 
macroscopic behaviors. Under positive feedback, the mutual reinforcement between the size of the giant component and local activation accelerates the growth of the giant component, 
leading to diverse transitions, including continuous, discontinuous, and double transitions. Delayed negative feedback effectively 
regulates the growth of the connectivity, causing the stable fixed point 
into stable oscillatory states. These oscillatory dynamics provide a robust 
theoretical foundation for understanding self-regulating systems in the real world.
We also found that the non-monotonic feedback function can cause chaotic behavior in the size of the giant component. 

Our finding suggests that the feedback in percolation is an important ingredient for modeling and understanding complex networked systems.
This study shows that the topology of a network is not a fixed property but a result of continuous self-adaptation, generating diverse behaviors, from explosive growth to stable oscillations and route to chaos.
These feedback processes are also manifested in adaptive network dynamics, reflecting a coevolutionary interplay between individual node behavior and the global network structure \cite{gross2006epidemic,vazquez2008generic,min2017fragmentation,scarpino2016effect,marceau2010adaptive}.
A direction for future work is to relate the feedback rules to link-rewiring processes~\cite{cho2025local,cho2026discontinuous} and to extend the framework to higher-order interactions~\cite{bianconi2021higher,battiston2026collective,sun2024higher}. 
In addition, since the route to chaos identified here relies on the map being unimodal, 
other functional forms of the feedback, such as multimodal ones, or 
multilayer settings~\cite{bianconi2018multilayer,sun2026triadic} in which the feedback becomes multidimensional may give rise to distinct universality classes and richer routes to chaos~\cite{sun2024higher,sun2026triadic}.

\section{\label{sec:Theory} Methods}

\subsection{The theoretical framework}
In this section, we derive self-consistency equations and their evolution 
for feedback percolation models on random networks with a degree distribution $P(k)$. 
We first consider a network where all links are initially inactive. In the initial step, each link is activated with probability $p$. 
Let $u_n$ be the probability that a node reached by following a randomly chosen link does not belong to the giant component at step $n$. 
We then define the degree generating function $G_0(x) = \sum_{k=0}^{\infty} P(k) x^k$. 
Since the feedback percolation model at step $n=0$ is equivalent to the ordinary bond percolation, defined on locally tree-like networks, $u_0$ is given by the generating function method \cite{newman2001random}, $u_0 = (1-p) + p G_1(u_0)$,
where $G_1(x) = G_0'(x) / G_0'(1)$ and $G_0'(x) = dG_0(x) / dx$
(see Supplementary Information for details). 
By solving it iteratively, we can obtain the value of $u_0$ at a fixed point. 
The size $S_0$ of the giant component is then given by 
$S_0 = 1 - G_0(u_0)$.

In the feedback percolation model, at each step $n$, the activation probability is determined by the size of the giant component at the previous step $S_{n-1}$, through the feedback function $f(S_{n-1})$. 
Combining the generating function formalism and the feedback effect, we arrive at the self-consistency equations for the feedback percolation model for $n \ge 1$ as:
\begin{align}
\label{Eq_un} u_{n} &= (1-p_{n}) + p_{n} G_1(u_{n}), \\
\label{Eq_sn} S_{n} &=  1 - G_0(u_{n}),
\end{align}
with the initial activation probability $p_0 = p$
and $u_0$ and $S_0$. 
By iteratively solving Eqs.~(\ref{Eq_pn}), (\ref{Eq_un}), and (\ref{Eq_sn}), we can obtain the size $S_n$ of the giant component at each step $n$.

\subsection{Stability Analysis}

Combining Eqs.~(\ref{Eq_pn}), (\ref{Eq_un}), and (\ref{Eq_sn}), we can obtain the implicit iterative map with a one-step delayed feedback:
\begin{align}
u_{n} &= 1-\big[p+f(1-G_0(u_{n-1})) \big] \big[1-G_1(u_{n}) \big].
\end{align}
To ensure that the system converges to a stable fixed point $u^*$, the magnitude of 
the derivative of the iterative map evaluated at that point satisfies 
the stability condition $\left| du_n / du_{n-1} \right|_{u^*} < 1$.
This condition ensures the dynamical stability of the system and guarantees that the iterative process successfully reaches the steady-state value $u^*$. 
If this criterion is not met, the system may exhibit unstable dynamical behavior.
This leads to the condition for the critical point (see the details in Supplementary Information):
\begin{align}
\label{Eq_stability}
\left| \frac{f'(1 - G_0(u_{n-1}))  [1 - G_1(u_n)] G_0'(u_{n-1})}{1 - [p + f(1 - G_0(u_{n-1}))] G_1'(u_n)} \right|_{u^*} =1,
\end{align}
where $f'(x)=df(x)/dx$.

Once the system converges to a stable fixed point $u^*$, the steady-state 
where $u_n = u_{n-1} = u^*$ is governed by the self-consistency equation:
$u^* = 1-\left[p+f(1-G_0(u^*)) \right] \left[1-G_1(u^*) \right]$.
According to the self-consistency equation, $u^*=1$ becomes a trivial solution corresponding to the non-percolating (NP) phase, $S^*=0$. 
The emergence of the giant component can be expected when the $u^* = 1$ solution becomes unstable. 
Defining $\phi(x)=1-\left[p+f(1-G_0(x)) \right] \left[1-G_1(x) \right]$, 
the condition leads to $\phi'(1)=1$ where $\phi'(x)=d\phi(x)/dx$.
Therefore, the percolation threshold $p_c$ for the percolation transition satisfies the following condition 
$p_c + f(0) = \frac{1}{G_1'(1)}$
where $G_1'(1) = (\langle k^2 \rangle - \langle k \rangle) / \langle k \rangle$. 
The value of $p_c$ is determined by both the network topology and the feedback function $f(S)$, rather than being fully determined by the degree distribution of the underlying networks as in ordinary percolation models.
Note that if $f(0)=0$, the percolation threshold reduces to the classical Molloy-Reed criterion.

Depending on the feedback functions and parameters, a non-trivial stable fixed point of $u^*$ can appear apart from the trivial solution $u^*=1$. 
The condition for the emergence of the stable point at $u^*$ is given by $\phi'(u^*)=1$, 
equivalent to the condition derived from Eq.~(\ref{Eq_stability}).
Specifically, the point $p_d$ of the emergence of the stable fixed point is given by: 
\begin{equation}
f'(S^*) [1 - G_1(u^*)] G_0'(u^*)=1 - [p_d + f(S^*)] G_1'(u^*).
\label{Eq_pd}
\end{equation}
At this point $p_d$, a stable fixed point can appear or disappear with a discontinuous jump.

Furthermore, when $d u_n / d u_{n-1} = -1$, the fixed point loses its stability, giving rise to an oscillatory behavior. We define the threshold for the onset of these stable oscillations as $p_o$. Specifically, $p_o$ can be expressed as
\begin{equation}
f'(S^*) [1 - G_1(u^*)] G_0'(u^*)= - 1 + [p_o + f(S^*)] G_1'(u^*).
\label{Eq_po}
\end{equation}
As the control parameter $p$ increases further, the system shows a sequence of period-doubling bifurcations. This cascade eventually leads the system to a chaotic regime, characterized by the aperiodic evolution of the size $S_n$ of the giant component.

\subsection{On ER graphs}
The generating functions for ER graphs are simply $G_0(x) = G_1(x) = e^{z(x-1)}$, thus the self-consistent equations for $u_n$ and $S_n$ in Eqs.~(\ref{Eq_un}) and (\ref{Eq_sn}) become effectively identical, leading to the relation $p_n S_n = 1 - u_n$. 
By substituting $p_n$ into the self-consistency equation for $S_n$, we arrive at the simplified equations:
\begin{align}    
p_n &= p + f(S_{n-1}),\\
S_{n} &= 1- \exp(- z S_n p_n).
\label{Eq_er}
\end{align}
The size $S_0$ of the giant component at the initial step $n=0$ is obtained from the self-consistent equation $S_0 = 1 - e^{-z S_0 p}$ with a bare probability $p$.
Note that the size $S_n$ depends on the size $S_{n-1}$ in the previous step by incorporating the feedback function $f(S_{n-1})$ in $p_n$.

\subsection{Numerical implementations}

In order to implement the feedback effect in the numerical simulation, we update the activation probability $p_n$ based on the value of $S_{n-1}$ from the previous time step. When $\Delta p = p_n-p_{n-1} >0 $, we additionally activate the inactive links with probability $\Delta p/(1-p_{n-1})$, thus increasing the overall activation probability to $p_n$. In contrast, when $\Delta p < 0$, each active link is deactivated with probability $(p_{n-1}-p_n)/p_{n-1}$. After this activation or deactivation step, we compute the size of the giant component $S_n$ and update the activation probability to obtain $p_{n+1}$. These procedures are repeated until the 
system arrives at the steady state, $\Delta p=0$ or equivalently $\Delta S = S_n - S_{n-1} =0$, 
whenever a stable fixed point exists.
When the dynamics does not converge to a fixed point, as in oscillatory or chaotic cases, we discard a transient period and analyze the long-time behavior of $S_n$.

\section*{Data availability}
The data supporting the findings of this work are available upon
request to the corresponding author.

\section*{Code availability}
The numerical simulation codes that support the findings of this study are available upon request to the corresponding author.

\bibliography{feedback}

\begin{thebibliography}{58}%
\makeatletter
\providecommand \@ifxundefined [1]{%
 \@ifx{#1\undefined}
}%
\providecommand \@ifnum [1]{%
 \ifnum #1\expandafter \@firstoftwo
 \else \expandafter \@secondoftwo
 \fi
}%
\providecommand \@ifx [1]{%
 \ifx #1\expandafter \@firstoftwo
 \else \expandafter \@secondoftwo
 \fi
}%
\providecommand \natexlab [1]{#1}%
\providecommand \enquote  [1]{``#1''}%
\providecommand \bibnamefont  [1]{#1}%
\providecommand \bibfnamefont [1]{#1}%
\providecommand \citenamefont [1]{#1}%
\providecommand \href@noop [0]{\@secondoftwo}%
\providecommand \href [0]{\begingroup \@sanitize@url \@href}%
\providecommand \@href[1]{\@@startlink{#1}\@@href}%
\providecommand \@@href[1]{\endgroup#1\@@endlink}%
\providecommand \@sanitize@url [0]{\catcode `\\12\catcode `\$12\catcode
  `\&12\catcode `\#12\catcode `\^12\catcode `\_12\catcode `\%12\relax}%
\providecommand \@@startlink[1]{}%
\providecommand \@@endlink[0]{}%
\providecommand \url  [0]{\begingroup\@sanitize@url \@url }%
\providecommand \@url [1]{\endgroup\@href {#1}{\urlprefix }}%
\providecommand \urlprefix  [0]{URL }%
\providecommand \Eprint [0]{\href }%
\providecommand \doibase [0]{https://doi.org/}%
\providecommand \selectlanguage [0]{\@gobble}%
\providecommand \bibinfo  [0]{\@secondoftwo}%
\providecommand \bibfield  [0]{\@secondoftwo}%
\providecommand \translation [1]{[#1]}%
\providecommand \BibitemOpen [0]{}%
\providecommand \bibitemStop [0]{}%
\providecommand \bibitemNoStop [0]{.\EOS\space}%
\providecommand \EOS [0]{\spacefactor3000\relax}%
\providecommand \BibitemShut  [1]{\csname bibitem#1\endcsname}%
\let\auto@bib@innerbib\@empty
\bibitem [{\citenamefont {Broadbent}\ and\ \citenamefont
  {Hammersley}(1957)}]{broadbent1957}%
  \BibitemOpen
  \bibfield  {author} {\bibinfo {author} {\bibfnamefont {S.~R.}\ \bibnamefont
  {Broadbent}}\ and\ \bibinfo {author} {\bibfnamefont {J.~M.}\ \bibnamefont
  {Hammersley}},\ }\bibfield  {title} {\bibinfo {title} {Percolation processes:
  I. crystals and mazes},\ }in\ \href@noop {} {\emph {\bibinfo {booktitle}
  {Mathematical proceedings of the Cambridge philosophical society}}},\
  Vol.~\bibinfo {volume} {53}\ (\bibinfo {organization} {Cambridge University
  Press},\ \bibinfo {year} {1957})\ pp.\ \bibinfo {pages}
  {629--641}\BibitemShut {NoStop}%
\bibitem [{\citenamefont {Stauffer}\ and\ \citenamefont
  {Aharony}(2018)}]{stauffer2018}%
  \BibitemOpen
  \bibfield  {author} {\bibinfo {author} {\bibfnamefont {D.}~\bibnamefont
  {Stauffer}}\ and\ \bibinfo {author} {\bibfnamefont {A.}~\bibnamefont
  {Aharony}},\ }\href@noop {} {\emph {\bibinfo {title} {Introduction to
  percolation theory}}}\ (\bibinfo  {publisher} {Taylor \& Francis},\ \bibinfo
  {year} {2018})\BibitemShut {NoStop}%
\bibitem [{\citenamefont {Kirkpatrick}(1973)}]{kirkpatrick1973}%
  \BibitemOpen
  \bibfield  {author} {\bibinfo {author} {\bibfnamefont {S.}~\bibnamefont
  {Kirkpatrick}},\ }\bibfield  {title} {\bibinfo {title} {Percolation and
  conduction},\ }\href@noop {} {\bibfield  {journal} {\bibinfo  {journal}
  {Reviews of modern physics}\ }\textbf {\bibinfo {volume} {45}},\ \bibinfo
  {pages} {574} (\bibinfo {year} {1973})}\BibitemShut {NoStop}%
\bibitem [{\citenamefont {Cardy}\ and\ \citenamefont
  {Grassberger}(1985)}]{cardy1985epidemic}%
  \BibitemOpen
  \bibfield  {author} {\bibinfo {author} {\bibfnamefont {J.~L.}\ \bibnamefont
  {Cardy}}\ and\ \bibinfo {author} {\bibfnamefont {P.}~\bibnamefont
  {Grassberger}},\ }\bibfield  {title} {\bibinfo {title} {Epidemic models and
  percolation},\ }\href@noop {} {\bibfield  {journal} {\bibinfo  {journal}
  {Journal of Physics A: Mathematical and General}\ }\textbf {\bibinfo {volume}
  {18}},\ \bibinfo {pages} {L267} (\bibinfo {year} {1985})}\BibitemShut
  {NoStop}%
\bibitem [{\citenamefont {Newman}(2002)}]{newman2002spread}%
  \BibitemOpen
  \bibfield  {author} {\bibinfo {author} {\bibfnamefont {M.}~\bibnamefont
  {Newman}},\ }\bibfield  {title} {\bibinfo {title} {Spread of epidemic disease
  on networks},\ }\href@noop {} {\bibfield  {journal} {\bibinfo  {journal}
  {Physical review E}\ }\textbf {\bibinfo {volume} {66}},\ \bibinfo {pages}
  {016128} (\bibinfo {year} {2002})}\BibitemShut {NoStop}%
\bibitem [{\citenamefont {Albert}\ \emph {et~al.}(2000)\citenamefont {Albert},
  \citenamefont {Jeong},\ and\ \citenamefont {Barab{\'a}si}}]{albert2000error}%
  \BibitemOpen
  \bibfield  {author} {\bibinfo {author} {\bibfnamefont {R.}~\bibnamefont
  {Albert}}, \bibinfo {author} {\bibfnamefont {H.}~\bibnamefont {Jeong}},\ and\
  \bibinfo {author} {\bibfnamefont {A.-L.}\ \bibnamefont {Barab{\'a}si}},\
  }\bibfield  {title} {\bibinfo {title} {Error and attack tolerance of complex
  networks},\ }\href@noop {} {\bibfield  {journal} {\bibinfo  {journal}
  {nature}\ }\textbf {\bibinfo {volume} {406}},\ \bibinfo {pages} {378}
  (\bibinfo {year} {2000})}\BibitemShut {NoStop}%
\bibitem [{\citenamefont {Buldyrev}\ \emph {et~al.}(2010)\citenamefont
  {Buldyrev}, \citenamefont {Parshani}, \citenamefont {Paul}, \citenamefont
  {Stanley},\ and\ \citenamefont {Havlin}}]{buldyrev2010catastrophic}%
  \BibitemOpen
  \bibfield  {author} {\bibinfo {author} {\bibfnamefont {S.~V.}\ \bibnamefont
  {Buldyrev}}, \bibinfo {author} {\bibfnamefont {R.}~\bibnamefont {Parshani}},
  \bibinfo {author} {\bibfnamefont {G.}~\bibnamefont {Paul}}, \bibinfo {author}
  {\bibfnamefont {H.~E.}\ \bibnamefont {Stanley}},\ and\ \bibinfo {author}
  {\bibfnamefont {S.}~\bibnamefont {Havlin}},\ }\bibfield  {title} {\bibinfo
  {title} {Catastrophic cascade of failures in interdependent networks},\
  }\href@noop {} {\bibfield  {journal} {\bibinfo  {journal} {Nature}\ }\textbf
  {\bibinfo {volume} {464}},\ \bibinfo {pages} {1025} (\bibinfo {year}
  {2010})}\BibitemShut {NoStop}%
\bibitem [{\citenamefont {Bonamassa}\ \emph {et~al.}(2025)\citenamefont
  {Bonamassa}, \citenamefont {Gross}, \citenamefont {Kert{\'e}sz},\ and\
  \citenamefont {Havlin}}]{bonamassa2025hybrid}%
  \BibitemOpen
  \bibfield  {author} {\bibinfo {author} {\bibfnamefont {I.}~\bibnamefont
  {Bonamassa}}, \bibinfo {author} {\bibfnamefont {B.}~\bibnamefont {Gross}},
  \bibinfo {author} {\bibfnamefont {J.}~\bibnamefont {Kert{\'e}sz}},\ and\
  \bibinfo {author} {\bibfnamefont {S.}~\bibnamefont {Havlin}},\ }\bibfield
  {title} {\bibinfo {title} {Hybrid universality classes of systemic
  cascades},\ }\href@noop {} {\bibfield  {journal} {\bibinfo  {journal} {Nature
  Communications}\ }\textbf {\bibinfo {volume} {16}},\ \bibinfo {pages} {1415}
  (\bibinfo {year} {2025})}\BibitemShut {NoStop}%
\bibitem [{\citenamefont {Artime}\ \emph {et~al.}(2024)\citenamefont {Artime},
  \citenamefont {Grassia}, \citenamefont {De~Domenico}, \citenamefont
  {Gleeson}, \citenamefont {Makse}, \citenamefont {Mangioni}, \citenamefont
  {Perc},\ and\ \citenamefont {Radicchi}}]{artime2024robustness}%
  \BibitemOpen
  \bibfield  {author} {\bibinfo {author} {\bibfnamefont {O.}~\bibnamefont
  {Artime}}, \bibinfo {author} {\bibfnamefont {M.}~\bibnamefont {Grassia}},
  \bibinfo {author} {\bibfnamefont {M.}~\bibnamefont {De~Domenico}}, \bibinfo
  {author} {\bibfnamefont {J.~P.}\ \bibnamefont {Gleeson}}, \bibinfo {author}
  {\bibfnamefont {H.~A.}\ \bibnamefont {Makse}}, \bibinfo {author}
  {\bibfnamefont {G.}~\bibnamefont {Mangioni}}, \bibinfo {author}
  {\bibfnamefont {M.}~\bibnamefont {Perc}},\ and\ \bibinfo {author}
  {\bibfnamefont {F.}~\bibnamefont {Radicchi}},\ }\bibfield  {title} {\bibinfo
  {title} {Robustness and resilience of complex networks},\ }\href@noop {}
  {\bibfield  {journal} {\bibinfo  {journal} {Nature Reviews Physics}\ }\textbf
  {\bibinfo {volume} {6}},\ \bibinfo {pages} {114} (\bibinfo {year}
  {2024})}\BibitemShut {NoStop}%
\bibitem [{\citenamefont {Meng}\ \emph {et~al.}(2023)\citenamefont {Meng},
  \citenamefont {Hu}, \citenamefont {Tian}, \citenamefont {Dong}, \citenamefont
  {Lambiotte}, \citenamefont {Gao},\ and\ \citenamefont
  {Havlin}}]{meng2023percolation}%
  \BibitemOpen
  \bibfield  {author} {\bibinfo {author} {\bibfnamefont {X.}~\bibnamefont
  {Meng}}, \bibinfo {author} {\bibfnamefont {X.}~\bibnamefont {Hu}}, \bibinfo
  {author} {\bibfnamefont {Y.}~\bibnamefont {Tian}}, \bibinfo {author}
  {\bibfnamefont {G.}~\bibnamefont {Dong}}, \bibinfo {author} {\bibfnamefont
  {R.}~\bibnamefont {Lambiotte}}, \bibinfo {author} {\bibfnamefont
  {J.}~\bibnamefont {Gao}},\ and\ \bibinfo {author} {\bibfnamefont
  {S.}~\bibnamefont {Havlin}},\ }\bibfield  {title} {\bibinfo {title}
  {Percolation theories for quantum networks},\ }\href@noop {} {\bibfield
  {journal} {\bibinfo  {journal} {Entropy}\ }\textbf {\bibinfo {volume} {25}},\
  \bibinfo {pages} {1564} (\bibinfo {year} {2023})}\BibitemShut {NoStop}%
\bibitem [{\citenamefont {Meng}\ \emph
  {et~al.}(2025{\natexlab{a}})\citenamefont {Meng}, \citenamefont {Hao},
  \citenamefont {R{\'a}th},\ and\ \citenamefont {Kov{\'a}cs}}]{meng2025path}%
  \BibitemOpen
  \bibfield  {author} {\bibinfo {author} {\bibfnamefont {X.}~\bibnamefont
  {Meng}}, \bibinfo {author} {\bibfnamefont {B.}~\bibnamefont {Hao}}, \bibinfo
  {author} {\bibfnamefont {B.}~\bibnamefont {R{\'a}th}},\ and\ \bibinfo
  {author} {\bibfnamefont {I.~A.}\ \bibnamefont {Kov{\'a}cs}},\ }\bibfield
  {title} {\bibinfo {title} {Path percolation in quantum communication
  networks},\ }\href@noop {} {\bibfield  {journal} {\bibinfo  {journal}
  {Physical review letters}\ }\textbf {\bibinfo {volume} {134}},\ \bibinfo
  {pages} {030803} (\bibinfo {year} {2025}{\natexlab{a}})}\BibitemShut
  {NoStop}%
\bibitem [{\citenamefont {Cirigliano}\ \emph {et~al.}(2026)\citenamefont
  {Cirigliano}, \citenamefont {Brosco}, \citenamefont {Castellano},
  \citenamefont {Felicetti}, \citenamefont {Pilozzi},\ and\ \citenamefont {van
  Heck}}]{cirigliano2026dynamical}%
  \BibitemOpen
  \bibfield  {author} {\bibinfo {author} {\bibfnamefont {L.}~\bibnamefont
  {Cirigliano}}, \bibinfo {author} {\bibfnamefont {V.}~\bibnamefont {Brosco}},
  \bibinfo {author} {\bibfnamefont {C.}~\bibnamefont {Castellano}}, \bibinfo
  {author} {\bibfnamefont {S.}~\bibnamefont {Felicetti}}, \bibinfo {author}
  {\bibfnamefont {L.}~\bibnamefont {Pilozzi}},\ and\ \bibinfo {author}
  {\bibfnamefont {B.}~\bibnamefont {van Heck}},\ }\bibfield  {title} {\bibinfo
  {title} {Dynamical entanglement percolation with spatially correlated
  disorder},\ }\href@noop {} {\bibfield  {journal} {\bibinfo  {journal} {arXiv
  preprint arXiv:2601.05925}\ } (\bibinfo {year} {2026})}\BibitemShut {NoStop}%
\bibitem [{\citenamefont {Dorogovtsev}\ \emph {et~al.}(2008)\citenamefont
  {Dorogovtsev}, \citenamefont {Goltsev},\ and\ \citenamefont
  {Mendes}}]{dorogovtsev2008critical}%
  \BibitemOpen
  \bibfield  {author} {\bibinfo {author} {\bibfnamefont {S.~N.}\ \bibnamefont
  {Dorogovtsev}}, \bibinfo {author} {\bibfnamefont {A.~V.}\ \bibnamefont
  {Goltsev}},\ and\ \bibinfo {author} {\bibfnamefont {J.~F.}\ \bibnamefont
  {Mendes}},\ }\bibfield  {title} {\bibinfo {title} {Critical phenomena in
  complex networks},\ }\href@noop {} {\bibfield  {journal} {\bibinfo  {journal}
  {Reviews of Modern Physics}\ }\textbf {\bibinfo {volume} {80}},\ \bibinfo
  {pages} {1275} (\bibinfo {year} {2008})}\BibitemShut {NoStop}%
\bibitem [{\citenamefont {Bianconi}(2018)}]{bianconi2018multilayer}%
  \BibitemOpen
  \bibfield  {author} {\bibinfo {author} {\bibfnamefont {G.}~\bibnamefont
  {Bianconi}},\ }\href@noop {} {\emph {\bibinfo {title} {Multilayer networks:
  structure and function}}}\ (\bibinfo  {publisher} {Oxford University Press},\
  \bibinfo {year} {2018})\BibitemShut {NoStop}%
\bibitem [{\citenamefont {Ara{\'u}jo}\ \emph {et~al.}(2014)\citenamefont
  {Ara{\'u}jo}, \citenamefont {Grassberger}, \citenamefont {Kahng},
  \citenamefont {Schrenk},\ and\ \citenamefont {Ziff}}]{araujo2014recent}%
  \BibitemOpen
  \bibfield  {author} {\bibinfo {author} {\bibfnamefont {N.}~\bibnamefont
  {Ara{\'u}jo}}, \bibinfo {author} {\bibfnamefont {P.}~\bibnamefont
  {Grassberger}}, \bibinfo {author} {\bibfnamefont {B.}~\bibnamefont {Kahng}},
  \bibinfo {author} {\bibfnamefont {K.}~\bibnamefont {Schrenk}},\ and\ \bibinfo
  {author} {\bibfnamefont {R.~M.}\ \bibnamefont {Ziff}},\ }\bibfield  {title}
  {\bibinfo {title} {Recent advances and open challenges in percolation},\
  }\href@noop {} {\bibfield  {journal} {\bibinfo  {journal} {The European
  Physical Journal Special Topics}\ }\textbf {\bibinfo {volume} {223}},\
  \bibinfo {pages} {2307} (\bibinfo {year} {2014})}\BibitemShut {NoStop}%
\bibitem [{\citenamefont {Lee}\ \emph {et~al.}(2018)\citenamefont {Lee},
  \citenamefont {Kahng}, \citenamefont {Cho}, \citenamefont {Goh},\ and\
  \citenamefont {Lee}}]{lee2018recent}%
  \BibitemOpen
  \bibfield  {author} {\bibinfo {author} {\bibfnamefont {D.}~\bibnamefont
  {Lee}}, \bibinfo {author} {\bibfnamefont {B.}~\bibnamefont {Kahng}}, \bibinfo
  {author} {\bibfnamefont {Y.}~\bibnamefont {Cho}}, \bibinfo {author}
  {\bibfnamefont {K.-I.}\ \bibnamefont {Goh}},\ and\ \bibinfo {author}
  {\bibfnamefont {D.-S.}\ \bibnamefont {Lee}},\ }\bibfield  {title} {\bibinfo
  {title} {Recent advances of percolation theory in complex networks},\
  }\href@noop {} {\bibfield  {journal} {\bibinfo  {journal} {Journal of the
  Korean Physical Society}\ }\textbf {\bibinfo {volume} {73}},\ \bibinfo
  {pages} {152} (\bibinfo {year} {2018})}\BibitemShut {NoStop}%
\bibitem [{\citenamefont {Cho}\ and\ \citenamefont
  {Kahng}(2026)}]{cho2026recent}%
  \BibitemOpen
  \bibfield  {author} {\bibinfo {author} {\bibfnamefont {Y.~S.}\ \bibnamefont
  {Cho}}\ and\ \bibinfo {author} {\bibfnamefont {B.}~\bibnamefont {Kahng}},\
  }\bibfield  {title} {\bibinfo {title} {Recent progress on hybrid percolation
  transitions},\ }\href@noop {} {\bibfield  {journal} {\bibinfo  {journal}
  {Entropy}\ }\textbf {\bibinfo {volume} {28}},\ \bibinfo {pages} {68}
  (\bibinfo {year} {2026})}\BibitemShut {NoStop}%
\bibitem [{\citenamefont {Christensen}\ and\ \citenamefont
  {Moloney}(2005)}]{christensen2005}%
  \BibitemOpen
  \bibfield  {author} {\bibinfo {author} {\bibfnamefont {K.}~\bibnamefont
  {Christensen}}\ and\ \bibinfo {author} {\bibfnamefont {N.~R.}\ \bibnamefont
  {Moloney}},\ }\href@noop {} {\emph {\bibinfo {title} {Complexity and
  criticality}}},\ Vol.~\bibinfo {volume} {1}\ (\bibinfo  {publisher} {Imperial
  College Press},\ \bibinfo {year} {2005})\BibitemShut {NoStop}%
\bibitem [{\citenamefont {Son}\ \emph {et~al.}(2012)\citenamefont {Son},
  \citenamefont {Bizhani}, \citenamefont {Christensen}, \citenamefont
  {Grassberger},\ and\ \citenamefont {Paczuski}}]{son2012percolation}%
  \BibitemOpen
  \bibfield  {author} {\bibinfo {author} {\bibfnamefont {S.-W.}\ \bibnamefont
  {Son}}, \bibinfo {author} {\bibfnamefont {G.}~\bibnamefont {Bizhani}},
  \bibinfo {author} {\bibfnamefont {C.}~\bibnamefont {Christensen}}, \bibinfo
  {author} {\bibfnamefont {P.}~\bibnamefont {Grassberger}},\ and\ \bibinfo
  {author} {\bibfnamefont {M.}~\bibnamefont {Paczuski}},\ }\bibfield  {title}
  {\bibinfo {title} {Percolation theory on interdependent networks based on
  epidemic spreading},\ }\href@noop {} {\bibfield  {journal} {\bibinfo
  {journal} {Europhysics Letters}\ }\textbf {\bibinfo {volume} {97}},\ \bibinfo
  {pages} {16006} (\bibinfo {year} {2012})}\BibitemShut {NoStop}%
\bibitem [{\citenamefont {Sun}\ \emph {et~al.}(2023)\citenamefont {Sun},
  \citenamefont {Radicchi}, \citenamefont {Kurths},\ and\ \citenamefont
  {Bianconi}}]{sun2023dynamic}%
  \BibitemOpen
  \bibfield  {author} {\bibinfo {author} {\bibfnamefont {H.}~\bibnamefont
  {Sun}}, \bibinfo {author} {\bibfnamefont {F.}~\bibnamefont {Radicchi}},
  \bibinfo {author} {\bibfnamefont {J.}~\bibnamefont {Kurths}},\ and\ \bibinfo
  {author} {\bibfnamefont {G.}~\bibnamefont {Bianconi}},\ }\bibfield  {title}
  {\bibinfo {title} {The dynamic nature of percolation on networks with triadic
  interactions},\ }\href@noop {} {\bibfield  {journal} {\bibinfo  {journal}
  {Nature Communications}\ }\textbf {\bibinfo {volume} {14}},\ \bibinfo {pages}
  {1308} (\bibinfo {year} {2023})}\BibitemShut {NoStop}%
\bibitem [{\citenamefont {Zhao}\ and\ \citenamefont
  {Bianconi}(2013)}]{zhao2013percolation}%
  \BibitemOpen
  \bibfield  {author} {\bibinfo {author} {\bibfnamefont {K.}~\bibnamefont
  {Zhao}}\ and\ \bibinfo {author} {\bibfnamefont {G.}~\bibnamefont
  {Bianconi}},\ }\bibfield  {title} {\bibinfo {title} {Percolation on
  interacting, antagonistic networks},\ }\href@noop {} {\bibfield  {journal}
  {\bibinfo  {journal} {Journal of Statistical Mechanics: Theory and
  Experiment}\ }\textbf {\bibinfo {volume} {2013}},\ \bibinfo {pages} {P05005}
  (\bibinfo {year} {2013})}\BibitemShut {NoStop}%
\bibitem [{\citenamefont {H{\'e}bert-Dufresne}\ \emph
  {et~al.}(2025)\citenamefont {H{\'e}bert-Dufresne}, \citenamefont {Lovato},
  \citenamefont {Burgio}, \citenamefont {Gleeson}, \citenamefont {Redner},\
  and\ \citenamefont {Krapivsky}}]{hebert2025self}%
  \BibitemOpen
  \bibfield  {author} {\bibinfo {author} {\bibfnamefont {L.}~\bibnamefont
  {H{\'e}bert-Dufresne}}, \bibinfo {author} {\bibfnamefont {J.}~\bibnamefont
  {Lovato}}, \bibinfo {author} {\bibfnamefont {G.}~\bibnamefont {Burgio}},
  \bibinfo {author} {\bibfnamefont {J.~P.}\ \bibnamefont {Gleeson}}, \bibinfo
  {author} {\bibfnamefont {S.}~\bibnamefont {Redner}},\ and\ \bibinfo {author}
  {\bibfnamefont {P.~L.}\ \bibnamefont {Krapivsky}},\ }\bibfield  {title}
  {\bibinfo {title} {Self-reinforcing cascades: A spreading model for beliefs
  or products of varying intensity or quality},\ }\href@noop {} {\bibfield
  {journal} {\bibinfo  {journal} {Physical Review Letters}\ }\textbf {\bibinfo
  {volume} {135}},\ \bibinfo {pages} {087401} (\bibinfo {year}
  {2025})}\BibitemShut {NoStop}%
\bibitem [{\citenamefont {Parshani}\ \emph {et~al.}(2010)\citenamefont
  {Parshani}, \citenamefont {Buldyrev},\ and\ \citenamefont
  {Havlin}}]{parshani2010interdependent}%
  \BibitemOpen
  \bibfield  {author} {\bibinfo {author} {\bibfnamefont {R.}~\bibnamefont
  {Parshani}}, \bibinfo {author} {\bibfnamefont {S.~V.}\ \bibnamefont
  {Buldyrev}},\ and\ \bibinfo {author} {\bibfnamefont {S.}~\bibnamefont
  {Havlin}},\ }\bibfield  {title} {\bibinfo {title} {Interdependent networks:
  Reducing the coupling strength leads to a change: from a first to second
  order percolation transition},\ }\href@noop {} {\bibfield  {journal}
  {\bibinfo  {journal} {Physical review letters}\ }\textbf {\bibinfo {volume}
  {105}},\ \bibinfo {pages} {048701} (\bibinfo {year} {2010})}\BibitemShut
  {NoStop}%
\bibitem [{\citenamefont {Baxter}\ \emph {et~al.}(2012)\citenamefont {Baxter},
  \citenamefont {Dorogovtsev}, \citenamefont {Goltsev},\ and\ \citenamefont
  {Mendes}}]{baxter2012avalanche}%
  \BibitemOpen
  \bibfield  {author} {\bibinfo {author} {\bibfnamefont {G.}~\bibnamefont
  {Baxter}}, \bibinfo {author} {\bibfnamefont {S.}~\bibnamefont {Dorogovtsev}},
  \bibinfo {author} {\bibfnamefont {A.}~\bibnamefont {Goltsev}},\ and\ \bibinfo
  {author} {\bibfnamefont {J.}~\bibnamefont {Mendes}},\ }\bibfield  {title}
  {\bibinfo {title} {Avalanche collapse of interdependent networks},\
  }\href@noop {} {\bibfield  {journal} {\bibinfo  {journal} {Physical review
  letters}\ }\textbf {\bibinfo {volume} {109}},\ \bibinfo {pages} {248701}
  (\bibinfo {year} {2012})}\BibitemShut {NoStop}%
\bibitem [{\citenamefont {Gao}\ \emph {et~al.}(2012)\citenamefont {Gao},
  \citenamefont {Buldyrev}, \citenamefont {Stanley},\ and\ \citenamefont
  {Havlin}}]{gao2012networks}%
  \BibitemOpen
  \bibfield  {author} {\bibinfo {author} {\bibfnamefont {J.}~\bibnamefont
  {Gao}}, \bibinfo {author} {\bibfnamefont {S.~V.}\ \bibnamefont {Buldyrev}},
  \bibinfo {author} {\bibfnamefont {H.~E.}\ \bibnamefont {Stanley}},\ and\
  \bibinfo {author} {\bibfnamefont {S.}~\bibnamefont {Havlin}},\ }\bibfield
  {title} {\bibinfo {title} {Networks formed from interdependent networks},\
  }\href@noop {} {\bibfield  {journal} {\bibinfo  {journal} {Nature physics}\
  }\textbf {\bibinfo {volume} {8}},\ \bibinfo {pages} {40} (\bibinfo {year}
  {2012})}\BibitemShut {NoStop}%
\bibitem [{\citenamefont {Zhou}\ \emph {et~al.}(2014)\citenamefont {Zhou},
  \citenamefont {Bashan}, \citenamefont {Cohen}, \citenamefont {Berezin},
  \citenamefont {Shnerb},\ and\ \citenamefont {Havlin}}]{zhou2014simultaneous}%
  \BibitemOpen
  \bibfield  {author} {\bibinfo {author} {\bibfnamefont {D.}~\bibnamefont
  {Zhou}}, \bibinfo {author} {\bibfnamefont {A.}~\bibnamefont {Bashan}},
  \bibinfo {author} {\bibfnamefont {R.}~\bibnamefont {Cohen}}, \bibinfo
  {author} {\bibfnamefont {Y.}~\bibnamefont {Berezin}}, \bibinfo {author}
  {\bibfnamefont {N.}~\bibnamefont {Shnerb}},\ and\ \bibinfo {author}
  {\bibfnamefont {S.}~\bibnamefont {Havlin}},\ }\bibfield  {title} {\bibinfo
  {title} {Simultaneous first-and second-order percolation transitions in
  interdependent networks},\ }\href@noop {} {\bibfield  {journal} {\bibinfo
  {journal} {Physical Review E}\ }\textbf {\bibinfo {volume} {90}},\ \bibinfo
  {pages} {012803} (\bibinfo {year} {2014})}\BibitemShut {NoStop}%
\bibitem [{\citenamefont {Bashan}\ \emph {et~al.}(2011)\citenamefont {Bashan},
  \citenamefont {Parshani},\ and\ \citenamefont
  {Havlin}}]{bashan2011percolation}%
  \BibitemOpen
  \bibfield  {author} {\bibinfo {author} {\bibfnamefont {A.}~\bibnamefont
  {Bashan}}, \bibinfo {author} {\bibfnamefont {R.}~\bibnamefont {Parshani}},\
  and\ \bibinfo {author} {\bibfnamefont {S.}~\bibnamefont {Havlin}},\
  }\bibfield  {title} {\bibinfo {title} {Percolation in networks composed of
  connectivity and dependency links},\ }\href@noop {} {\bibfield  {journal}
  {\bibinfo  {journal} {Physical Review E—Statistical, Nonlinear, and Soft
  Matter Physics}\ }\textbf {\bibinfo {volume} {83}},\ \bibinfo {pages}
  {051127} (\bibinfo {year} {2011})}\BibitemShut {NoStop}%
\bibitem [{\citenamefont {Kuehn}\ and\ \citenamefont
  {Bick}(2021)}]{kuehn2021universal}%
  \BibitemOpen
  \bibfield  {author} {\bibinfo {author} {\bibfnamefont {C.}~\bibnamefont
  {Kuehn}}\ and\ \bibinfo {author} {\bibfnamefont {C.}~\bibnamefont {Bick}},\
  }\bibfield  {title} {\bibinfo {title} {A universal route to explosive
  phenomena},\ }\href@noop {} {\bibfield  {journal} {\bibinfo  {journal}
  {Science advances}\ }\textbf {\bibinfo {volume} {7}},\ \bibinfo {pages}
  {eabe3824} (\bibinfo {year} {2021})}\BibitemShut {NoStop}%
\bibitem [{\citenamefont {Mill{\'a}n}\ \emph {et~al.}(2025)\citenamefont
  {Mill{\'a}n}, \citenamefont {Sun}, \citenamefont {Giambagli}, \citenamefont
  {Muolo}, \citenamefont {Carletti}, \citenamefont {Torres}, \citenamefont
  {Radicchi}, \citenamefont {Kurths},\ and\ \citenamefont
  {Bianconi}}]{millan2025topology}%
  \BibitemOpen
  \bibfield  {author} {\bibinfo {author} {\bibfnamefont {A.~P.}\ \bibnamefont
  {Mill{\'a}n}}, \bibinfo {author} {\bibfnamefont {H.}~\bibnamefont {Sun}},
  \bibinfo {author} {\bibfnamefont {L.}~\bibnamefont {Giambagli}}, \bibinfo
  {author} {\bibfnamefont {R.}~\bibnamefont {Muolo}}, \bibinfo {author}
  {\bibfnamefont {T.}~\bibnamefont {Carletti}}, \bibinfo {author}
  {\bibfnamefont {J.~J.}\ \bibnamefont {Torres}}, \bibinfo {author}
  {\bibfnamefont {F.}~\bibnamefont {Radicchi}}, \bibinfo {author}
  {\bibfnamefont {J.}~\bibnamefont {Kurths}},\ and\ \bibinfo {author}
  {\bibfnamefont {G.}~\bibnamefont {Bianconi}},\ }\bibfield  {title} {\bibinfo
  {title} {Topology shapes dynamics of higher-order networks},\ }\href@noop {}
  {\bibfield  {journal} {\bibinfo  {journal} {Nature Physics}\ }\textbf
  {\bibinfo {volume} {21}},\ \bibinfo {pages} {353} (\bibinfo {year}
  {2025})}\BibitemShut {NoStop}%
\bibitem [{\citenamefont {Sun}\ and\ \citenamefont
  {Bianconi}(2024)}]{sun2024higher}%
  \BibitemOpen
  \bibfield  {author} {\bibinfo {author} {\bibfnamefont {H.}~\bibnamefont
  {Sun}}\ and\ \bibinfo {author} {\bibfnamefont {G.}~\bibnamefont {Bianconi}},\
  }\bibfield  {title} {\bibinfo {title} {Higher-order triadic percolation on
  random hypergraphs},\ }\href@noop {} {\bibfield  {journal} {\bibinfo
  {journal} {Physical Review E}\ }\textbf {\bibinfo {volume} {110}},\ \bibinfo
  {pages} {064315} (\bibinfo {year} {2024})}\BibitemShut {NoStop}%
\bibitem [{\citenamefont {Sun}\ \emph {et~al.}(2026)\citenamefont {Sun},
  \citenamefont {Radicchi},\ and\ \citenamefont {Bianconi}}]{sun2026triadic}%
  \BibitemOpen
  \bibfield  {author} {\bibinfo {author} {\bibfnamefont {H.}~\bibnamefont
  {Sun}}, \bibinfo {author} {\bibfnamefont {F.}~\bibnamefont {Radicchi}},\ and\
  \bibinfo {author} {\bibfnamefont {G.}~\bibnamefont {Bianconi}},\ }\bibfield
  {title} {\bibinfo {title} {Triadic percolation on multilayer networks},\
  }\href@noop {} {\bibfield  {journal} {\bibinfo  {journal} {Physical Review
  E}\ }\textbf {\bibinfo {volume} {113}},\ \bibinfo {pages} {014313} (\bibinfo
  {year} {2026})}\BibitemShut {NoStop}%
\bibitem [{\citenamefont {Mill{\'a}n}\ \emph {et~al.}(2024)\citenamefont
  {Mill{\'a}n}, \citenamefont {Sun}, \citenamefont {Torres},\ and\
  \citenamefont {Bianconi}}]{millan2024triadic}%
  \BibitemOpen
  \bibfield  {author} {\bibinfo {author} {\bibfnamefont {A.~P.}\ \bibnamefont
  {Mill{\'a}n}}, \bibinfo {author} {\bibfnamefont {H.}~\bibnamefont {Sun}},
  \bibinfo {author} {\bibfnamefont {J.~J.}\ \bibnamefont {Torres}},\ and\
  \bibinfo {author} {\bibfnamefont {G.}~\bibnamefont {Bianconi}},\ }\bibfield
  {title} {\bibinfo {title} {Triadic percolation induces dynamical topological
  patterns in higher-order networks},\ }\href@noop {} {\bibfield  {journal}
  {\bibinfo  {journal} {PNAS Nexus}\ }\textbf {\bibinfo {volume} {3}},\
  \bibinfo {pages} {pgae270} (\bibinfo {year} {2024})}\BibitemShut {NoStop}%
\bibitem [{\citenamefont {Aghaei}\ \emph {et~al.}(2026)\citenamefont {Aghaei},
  \citenamefont {Saberi}, \citenamefont {Kantz},\ and\ \citenamefont
  {Kurths}}]{aghaei2026superstable}%
  \BibitemOpen
  \bibfield  {author} {\bibinfo {author} {\bibfnamefont {F.}~\bibnamefont
  {Aghaei}}, \bibinfo {author} {\bibfnamefont {A.~A.}\ \bibnamefont {Saberi}},
  \bibinfo {author} {\bibfnamefont {H.}~\bibnamefont {Kantz}},\ and\ \bibinfo
  {author} {\bibfnamefont {J.}~\bibnamefont {Kurths}},\ }\bibfield  {title}
  {\bibinfo {title} {Superstable geometry in triadic percolation},\ }\href@noop
  {} {\bibfield  {journal} {\bibinfo  {journal} {Physical Review E}\ }\textbf
  {\bibinfo {volume} {113}},\ \bibinfo {pages} {024306} (\bibinfo {year}
  {2026})}\BibitemShut {NoStop}%
\bibitem [{\citenamefont {Achlioptas}\ \emph {et~al.}(2009)\citenamefont
  {Achlioptas}, \citenamefont {D'souza},\ and\ \citenamefont
  {Spencer}}]{achlioptas2009explosive}%
  \BibitemOpen
  \bibfield  {author} {\bibinfo {author} {\bibfnamefont {D.}~\bibnamefont
  {Achlioptas}}, \bibinfo {author} {\bibfnamefont {R.~M.}\ \bibnamefont
  {D'souza}},\ and\ \bibinfo {author} {\bibfnamefont {J.}~\bibnamefont
  {Spencer}},\ }\bibfield  {title} {\bibinfo {title} {Explosive percolation in
  random networks},\ }\href@noop {} {\bibfield  {journal} {\bibinfo  {journal}
  {science}\ }\textbf {\bibinfo {volume} {323}},\ \bibinfo {pages} {1453}
  (\bibinfo {year} {2009})}\BibitemShut {NoStop}%
\bibitem [{\citenamefont {D'Souza}\ \emph {et~al.}(2019)\citenamefont
  {D'Souza}, \citenamefont {G{\'o}mez-Gardenes}, \citenamefont {Nagler},\ and\
  \citenamefont {Arenas}}]{dsouza2019explosive}%
  \BibitemOpen
  \bibfield  {author} {\bibinfo {author} {\bibfnamefont {R.~M.}\ \bibnamefont
  {D'Souza}}, \bibinfo {author} {\bibfnamefont {J.}~\bibnamefont
  {G{\'o}mez-Gardenes}}, \bibinfo {author} {\bibfnamefont {J.}~\bibnamefont
  {Nagler}},\ and\ \bibinfo {author} {\bibfnamefont {A.}~\bibnamefont
  {Arenas}},\ }\bibfield  {title} {\bibinfo {title} {Explosive phenomena in
  complex networks},\ }\href@noop {} {\bibfield  {journal} {\bibinfo  {journal}
  {Advances in Physics}\ }\textbf {\bibinfo {volume} {68}},\ \bibinfo {pages}
  {123} (\bibinfo {year} {2019})}\BibitemShut {NoStop}%
\bibitem [{\citenamefont {Danziger}\ \emph {et~al.}(2019)\citenamefont
  {Danziger}, \citenamefont {Bonamassa}, \citenamefont {Boccaletti},\ and\
  \citenamefont {Havlin}}]{danziger2019dynamic}%
  \BibitemOpen
  \bibfield  {author} {\bibinfo {author} {\bibfnamefont {M.~M.}\ \bibnamefont
  {Danziger}}, \bibinfo {author} {\bibfnamefont {I.}~\bibnamefont {Bonamassa}},
  \bibinfo {author} {\bibfnamefont {S.}~\bibnamefont {Boccaletti}},\ and\
  \bibinfo {author} {\bibfnamefont {S.}~\bibnamefont {Havlin}},\ }\bibfield
  {title} {\bibinfo {title} {Dynamic interdependence and competition in
  multilayer networks},\ }\href@noop {} {\bibfield  {journal} {\bibinfo
  {journal} {Nature Physics}\ }\textbf {\bibinfo {volume} {15}},\ \bibinfo
  {pages} {178} (\bibinfo {year} {2019})}\BibitemShut {NoStop}%
\bibitem [{\citenamefont {Bonamassa}\ \emph {et~al.}(2021)\citenamefont
  {Bonamassa}, \citenamefont {Gross},\ and\ \citenamefont
  {Havlin}}]{bonamassa2021realizing}%
  \BibitemOpen
  \bibfield  {author} {\bibinfo {author} {\bibfnamefont {I.}~\bibnamefont
  {Bonamassa}}, \bibinfo {author} {\bibfnamefont {B.}~\bibnamefont {Gross}},\
  and\ \bibinfo {author} {\bibfnamefont {S.}~\bibnamefont {Havlin}},\
  }\bibfield  {title} {\bibinfo {title} {Realizing interdependent couplings as
  thermal or higher-order interactions},\ }\href@noop {} {\bibfield  {journal}
  {\bibinfo  {journal} {arXiv preprint arXiv:2110.08907}\ } (\bibinfo {year}
  {2021})}\BibitemShut {NoStop}%
\bibitem [{\citenamefont {Gross}\ \emph {et~al.}(2024)\citenamefont {Gross},
  \citenamefont {Bonamassa},\ and\ \citenamefont
  {Havlin}}]{gross2024microscopic}%
  \BibitemOpen
  \bibfield  {author} {\bibinfo {author} {\bibfnamefont {B.}~\bibnamefont
  {Gross}}, \bibinfo {author} {\bibfnamefont {I.}~\bibnamefont {Bonamassa}},\
  and\ \bibinfo {author} {\bibfnamefont {S.}~\bibnamefont {Havlin}},\
  }\bibfield  {title} {\bibinfo {title} {Microscopic intervention yields abrupt
  transition in interdependent ferromagnetic networks},\ }\href@noop {}
  {\bibfield  {journal} {\bibinfo  {journal} {Physical Review Letters}\
  }\textbf {\bibinfo {volume} {132}},\ \bibinfo {pages} {227401} (\bibinfo
  {year} {2024})}\BibitemShut {NoStop}%
\bibitem [{\citenamefont {Bonamassa}\ \emph {et~al.}(2023)\citenamefont
  {Bonamassa}, \citenamefont {Gross}, \citenamefont {Laav}, \citenamefont
  {Volotsenko}, \citenamefont {Frydman},\ and\ \citenamefont
  {Havlin}}]{bonamassa2023interdependent}%
  \BibitemOpen
  \bibfield  {author} {\bibinfo {author} {\bibfnamefont {I.}~\bibnamefont
  {Bonamassa}}, \bibinfo {author} {\bibfnamefont {B.}~\bibnamefont {Gross}},
  \bibinfo {author} {\bibfnamefont {M.}~\bibnamefont {Laav}}, \bibinfo {author}
  {\bibfnamefont {I.}~\bibnamefont {Volotsenko}}, \bibinfo {author}
  {\bibfnamefont {A.}~\bibnamefont {Frydman}},\ and\ \bibinfo {author}
  {\bibfnamefont {S.}~\bibnamefont {Havlin}},\ }\bibfield  {title} {\bibinfo
  {title} {Interdependent superconducting networks},\ }\href@noop {} {\bibfield
   {journal} {\bibinfo  {journal} {Nature Physics}\ }\textbf {\bibinfo {volume}
  {19}},\ \bibinfo {pages} {1163} (\bibinfo {year} {2023})}\BibitemShut
  {NoStop}%
\bibitem [{\citenamefont {Ma}\ \emph {et~al.}(2025{\natexlab{a}})\citenamefont
  {Ma}, \citenamefont {Sudakow}, \citenamefont {Krapivsky},\ and\ \citenamefont
  {Vakulenko}}]{ma2025dynamics}%
  \BibitemOpen
  \bibfield  {author} {\bibinfo {author} {\bibfnamefont {Y.-P.}\ \bibnamefont
  {Ma}}, \bibinfo {author} {\bibfnamefont {I.}~\bibnamefont {Sudakow}},
  \bibinfo {author} {\bibfnamefont {P.}~\bibnamefont {Krapivsky}},\ and\
  \bibinfo {author} {\bibfnamefont {S.~A.}\ \bibnamefont {Vakulenko}},\
  }\bibfield  {title} {\bibinfo {title} {Dynamics of feedback ising model},\
  }\href@noop {} {\bibfield  {journal} {\bibinfo  {journal} {arXiv preprint
  arXiv:2510.07301}\ } (\bibinfo {year} {2025}{\natexlab{a}})}\BibitemShut
  {NoStop}%
\bibitem [{\citenamefont {Ma}\ \emph {et~al.}(2025{\natexlab{b}})\citenamefont
  {Ma}, \citenamefont {Sudakow},\ and\ \citenamefont
  {Krapivsky}}]{ma2025mixed}%
  \BibitemOpen
  \bibfield  {author} {\bibinfo {author} {\bibfnamefont {Y.-P.}\ \bibnamefont
  {Ma}}, \bibinfo {author} {\bibfnamefont {I.}~\bibnamefont {Sudakow}},\ and\
  \bibinfo {author} {\bibfnamefont {P.}~\bibnamefont {Krapivsky}},\ }\bibfield
  {title} {\bibinfo {title} {Mixed phases in feedback ising models},\
  }\href@noop {} {\bibfield  {journal} {\bibinfo  {journal} {arXiv preprint
  arXiv:2506.09025}\ } (\bibinfo {year} {2025}{\natexlab{b}})}\BibitemShut
  {NoStop}%
\bibitem [{\citenamefont {Meng}\ \emph
  {et~al.}(2025{\natexlab{b}})\citenamefont {Meng}, \citenamefont {Piparo},
  \citenamefont {Nemoto},\ and\ \citenamefont {Kov{\'a}cs}}]{meng2025quantum}%
  \BibitemOpen
  \bibfield  {author} {\bibinfo {author} {\bibfnamefont {X.}~\bibnamefont
  {Meng}}, \bibinfo {author} {\bibfnamefont {N.~L.}\ \bibnamefont {Piparo}},
  \bibinfo {author} {\bibfnamefont {K.}~\bibnamefont {Nemoto}},\ and\ \bibinfo
  {author} {\bibfnamefont {I.~A.}\ \bibnamefont {Kov{\'a}cs}},\ }\bibfield
  {title} {\bibinfo {title} {Quantum communication networks enhanced by
  distributed quantum memories},\ }\href@noop {} {\bibfield  {journal}
  {\bibinfo  {journal} {Quantum}\ }\textbf {\bibinfo {volume} {9}},\ \bibinfo
  {pages} {1948} (\bibinfo {year} {2025}{\natexlab{b}})}\BibitemShut {NoStop}%
\bibitem [{\citenamefont {Funk}\ \emph {et~al.}(2009)\citenamefont {Funk},
  \citenamefont {Gilad}, \citenamefont {Watkins},\ and\ \citenamefont
  {Jansen}}]{funk2009spread}%
  \BibitemOpen
  \bibfield  {author} {\bibinfo {author} {\bibfnamefont {S.}~\bibnamefont
  {Funk}}, \bibinfo {author} {\bibfnamefont {E.}~\bibnamefont {Gilad}},
  \bibinfo {author} {\bibfnamefont {C.}~\bibnamefont {Watkins}},\ and\ \bibinfo
  {author} {\bibfnamefont {V.~A.}\ \bibnamefont {Jansen}},\ }\bibfield  {title}
  {\bibinfo {title} {The spread of awareness and its impact on epidemic
  outbreaks},\ }\href@noop {} {\bibfield  {journal} {\bibinfo  {journal}
  {Proceedings of the National Academy of Sciences}\ }\textbf {\bibinfo
  {volume} {106}},\ \bibinfo {pages} {6872} (\bibinfo {year}
  {2009})}\BibitemShut {NoStop}%
\bibitem [{\citenamefont {Hebb}(2005)}]{hebb2005}%
  \BibitemOpen
  \bibfield  {author} {\bibinfo {author} {\bibfnamefont {D.~O.}\ \bibnamefont
  {Hebb}},\ }\href@noop {} {\emph {\bibinfo {title} {The organization of
  behavior: A neuropsychological theory}}}\ (\bibinfo  {publisher} {Psychology
  press},\ \bibinfo {year} {2005})\BibitemShut {NoStop}%
\bibitem [{\citenamefont {Rapisardi}\ \emph {et~al.}(2022)\citenamefont
  {Rapisardi}, \citenamefont {Kryven},\ and\ \citenamefont
  {Arenas}}]{rapisardi2022}%
  \BibitemOpen
  \bibfield  {author} {\bibinfo {author} {\bibfnamefont {G.}~\bibnamefont
  {Rapisardi}}, \bibinfo {author} {\bibfnamefont {I.}~\bibnamefont {Kryven}},\
  and\ \bibinfo {author} {\bibfnamefont {A.}~\bibnamefont {Arenas}},\
  }\bibfield  {title} {\bibinfo {title} {Percolation in networks with local
  homeostatic plasticity},\ }\href@noop {} {\bibfield  {journal} {\bibinfo
  {journal} {Nature Communications}\ }\textbf {\bibinfo {volume} {13}},\
  \bibinfo {pages} {122} (\bibinfo {year} {2022})}\BibitemShut {NoStop}%
\bibitem [{\citenamefont {Strogatz}(2024)}]{strogatz2024nonlinear}%
  \BibitemOpen
  \bibfield  {author} {\bibinfo {author} {\bibfnamefont {S.~H.}\ \bibnamefont
  {Strogatz}},\ }\href@noop {} {\emph {\bibinfo {title} {Nonlinear dynamics and
  chaos: with applications to physics, biology, chemistry, and engineering}}}\
  (\bibinfo  {publisher} {Chapman and Hall/CRC},\ \bibinfo {year}
  {2024})\BibitemShut {NoStop}%
\bibitem [{\citenamefont {Feigenbaum}(1978)}]{feigenbaum1978quantitative}%
  \BibitemOpen
  \bibfield  {author} {\bibinfo {author} {\bibfnamefont {M.~J.}\ \bibnamefont
  {Feigenbaum}},\ }\bibfield  {title} {\bibinfo {title} {Quantitative
  universality for a class of nonlinear transformations},\ }\href@noop {}
  {\bibfield  {journal} {\bibinfo  {journal} {Journal of statistical physics}\
  }\textbf {\bibinfo {volume} {19}},\ \bibinfo {pages} {25} (\bibinfo {year}
  {1978})}\BibitemShut {NoStop}%
\bibitem [{\citenamefont {Cho}\ \emph {et~al.}(2013)\citenamefont {Cho},
  \citenamefont {Hwang}, \citenamefont {Herrmann},\ and\ \citenamefont
  {Kahng}}]{cho2013avoiding}%
  \BibitemOpen
  \bibfield  {author} {\bibinfo {author} {\bibfnamefont {Y.~S.}\ \bibnamefont
  {Cho}}, \bibinfo {author} {\bibfnamefont {S.}~\bibnamefont {Hwang}}, \bibinfo
  {author} {\bibfnamefont {H.~J.}\ \bibnamefont {Herrmann}},\ and\ \bibinfo
  {author} {\bibfnamefont {B.}~\bibnamefont {Kahng}},\ }\bibfield  {title}
  {\bibinfo {title} {Avoiding a spanning cluster in percolation models},\
  }\href@noop {} {\bibfield  {journal} {\bibinfo  {journal} {Science}\ }\textbf
  {\bibinfo {volume} {339}},\ \bibinfo {pages} {1185} (\bibinfo {year}
  {2013})}\BibitemShut {NoStop}%
\bibitem [{\citenamefont {Gross}\ \emph {et~al.}(2006)\citenamefont {Gross},
  \citenamefont {D’Lima},\ and\ \citenamefont {Blasius}}]{gross2006epidemic}%
  \BibitemOpen
  \bibfield  {author} {\bibinfo {author} {\bibfnamefont {T.}~\bibnamefont
  {Gross}}, \bibinfo {author} {\bibfnamefont {C.~J.~D.}\ \bibnamefont
  {D’Lima}},\ and\ \bibinfo {author} {\bibfnamefont {B.}~\bibnamefont
  {Blasius}},\ }\bibfield  {title} {\bibinfo {title} {Epidemic dynamics on an
  adaptive network},\ }\href@noop {} {\bibfield  {journal} {\bibinfo  {journal}
  {Physical review letters}\ }\textbf {\bibinfo {volume} {96}},\ \bibinfo
  {pages} {208701} (\bibinfo {year} {2006})}\BibitemShut {NoStop}%
\bibitem [{\citenamefont {Vazquez}\ \emph {et~al.}(2008)\citenamefont
  {Vazquez}, \citenamefont {Egu{\'\i}luz},\ and\ \citenamefont
  {Miguel}}]{vazquez2008generic}%
  \BibitemOpen
  \bibfield  {author} {\bibinfo {author} {\bibfnamefont {F.}~\bibnamefont
  {Vazquez}}, \bibinfo {author} {\bibfnamefont {V.~M.}\ \bibnamefont
  {Egu{\'\i}luz}},\ and\ \bibinfo {author} {\bibfnamefont {M.~S.}\ \bibnamefont
  {Miguel}},\ }\bibfield  {title} {\bibinfo {title} {Generic absorbing
  transition in coevolution dynamics},\ }\href@noop {} {\bibfield  {journal}
  {\bibinfo  {journal} {Physical review letters}\ }\textbf {\bibinfo {volume}
  {100}},\ \bibinfo {pages} {108702} (\bibinfo {year} {2008})}\BibitemShut
  {NoStop}%
\bibitem [{\citenamefont {Min}\ and\ \citenamefont
  {San~Miguel}(2017)}]{min2017fragmentation}%
  \BibitemOpen
  \bibfield  {author} {\bibinfo {author} {\bibfnamefont {B.}~\bibnamefont
  {Min}}\ and\ \bibinfo {author} {\bibfnamefont {M.}~\bibnamefont
  {San~Miguel}},\ }\bibfield  {title} {\bibinfo {title} {Fragmentation
  transitions in a coevolving nonlinear voter model},\ }\href@noop {}
  {\bibfield  {journal} {\bibinfo  {journal} {Scientific reports}\ }\textbf
  {\bibinfo {volume} {7}},\ \bibinfo {pages} {12864} (\bibinfo {year}
  {2017})}\BibitemShut {NoStop}%
\bibitem [{\citenamefont {Scarpino}\ \emph {et~al.}(2016)\citenamefont
  {Scarpino}, \citenamefont {Allard},\ and\ \citenamefont
  {H{\'e}bert-Dufresne}}]{scarpino2016effect}%
  \BibitemOpen
  \bibfield  {author} {\bibinfo {author} {\bibfnamefont {S.~V.}\ \bibnamefont
  {Scarpino}}, \bibinfo {author} {\bibfnamefont {A.}~\bibnamefont {Allard}},\
  and\ \bibinfo {author} {\bibfnamefont {L.}~\bibnamefont
  {H{\'e}bert-Dufresne}},\ }\bibfield  {title} {\bibinfo {title} {The effect of
  a prudent adaptive behaviour on disease transmission},\ }\href@noop {}
  {\bibfield  {journal} {\bibinfo  {journal} {Nature Physics}\ }\textbf
  {\bibinfo {volume} {12}},\ \bibinfo {pages} {1042} (\bibinfo {year}
  {2016})}\BibitemShut {NoStop}%
\bibitem [{\citenamefont {Marceau}\ \emph {et~al.}(2010)\citenamefont
  {Marceau}, \citenamefont {No{\"e}l}, \citenamefont {H{\'e}bert-Dufresne},
  \citenamefont {Allard},\ and\ \citenamefont
  {Dub{\'e}}}]{marceau2010adaptive}%
  \BibitemOpen
  \bibfield  {author} {\bibinfo {author} {\bibfnamefont {V.}~\bibnamefont
  {Marceau}}, \bibinfo {author} {\bibfnamefont {P.-A.}\ \bibnamefont
  {No{\"e}l}}, \bibinfo {author} {\bibfnamefont {L.}~\bibnamefont
  {H{\'e}bert-Dufresne}}, \bibinfo {author} {\bibfnamefont {A.}~\bibnamefont
  {Allard}},\ and\ \bibinfo {author} {\bibfnamefont {L.~J.}\ \bibnamefont
  {Dub{\'e}}},\ }\bibfield  {title} {\bibinfo {title} {Adaptive networks:
  Coevolution of disease and topology},\ }\href@noop {} {\bibfield  {journal}
  {\bibinfo  {journal} {Physical Review E—Statistical, Nonlinear, and Soft
  Matter Physics}\ }\textbf {\bibinfo {volume} {82}},\ \bibinfo {pages}
  {036116} (\bibinfo {year} {2010})}\BibitemShut {NoStop}%
\bibitem [{\citenamefont {Cho}(2025)}]{cho2025local}%
  \BibitemOpen
  \bibfield  {author} {\bibinfo {author} {\bibfnamefont {Y.~S.}\ \bibnamefont
  {Cho}},\ }\bibfield  {title} {\bibinfo {title} {Local suppression by link
  rewiring reveals discontinuous percolation transitions},\ }\href@noop {}
  {\bibfield  {journal} {\bibinfo  {journal} {Science Advances}\ }\textbf
  {\bibinfo {volume} {11}},\ \bibinfo {pages} {eadt0355} (\bibinfo {year}
  {2025})}\BibitemShut {NoStop}%
\bibitem [{\citenamefont {Cho}(2026)}]{cho2026discontinuous}%
  \BibitemOpen
  \bibfield  {author} {\bibinfo {author} {\bibfnamefont {Y.~S.}\ \bibnamefont
  {Cho}},\ }\bibfield  {title} {\bibinfo {title} {Discontinuous transition in
  explosive percolation via local suppression},\ }\href@noop {} {\bibfield
  {journal} {\bibinfo  {journal} {Physical Review E}\ }\textbf {\bibinfo
  {volume} {113}},\ \bibinfo {pages} {034307} (\bibinfo {year}
  {2026})}\BibitemShut {NoStop}%
\bibitem [{\citenamefont {Bianconi}(2021)}]{bianconi2021higher}%
  \BibitemOpen
  \bibfield  {author} {\bibinfo {author} {\bibfnamefont {G.}~\bibnamefont
  {Bianconi}},\ }\href@noop {} {\emph {\bibinfo {title} {Higher-order
  networks}}}\ (\bibinfo  {publisher} {Cambridge University Press},\ \bibinfo
  {year} {2021})\BibitemShut {NoStop}%
\bibitem [{\citenamefont {Battiston}\ \emph {et~al.}(2026)\citenamefont
  {Battiston}, \citenamefont {Bick}, \citenamefont {Lucas}, \citenamefont
  {Mill{\'a}n}, \citenamefont {Skardal},\ and\ \citenamefont
  {Zhang}}]{battiston2026collective}%
  \BibitemOpen
  \bibfield  {author} {\bibinfo {author} {\bibfnamefont {F.}~\bibnamefont
  {Battiston}}, \bibinfo {author} {\bibfnamefont {C.}~\bibnamefont {Bick}},
  \bibinfo {author} {\bibfnamefont {M.}~\bibnamefont {Lucas}}, \bibinfo
  {author} {\bibfnamefont {A.~P.}\ \bibnamefont {Mill{\'a}n}}, \bibinfo
  {author} {\bibfnamefont {P.~S.}\ \bibnamefont {Skardal}},\ and\ \bibinfo
  {author} {\bibfnamefont {Y.}~\bibnamefont {Zhang}},\ }\bibfield  {title}
  {\bibinfo {title} {Collective dynamics on higher-order networks},\
  }\href@noop {} {\bibfield  {journal} {\bibinfo  {journal} {Nature Reviews
  Physics}\ ,\ \bibinfo {pages} {1}} (\bibinfo {year} {2026})}\BibitemShut
  {NoStop}%
\bibitem [{\citenamefont {Newman}\ \emph {et~al.}(2001)\citenamefont {Newman},
  \citenamefont {Strogatz},\ and\ \citenamefont {Watts}}]{newman2001random}%
  \BibitemOpen
  \bibfield  {author} {\bibinfo {author} {\bibfnamefont {M.~E.}\ \bibnamefont
  {Newman}}, \bibinfo {author} {\bibfnamefont {S.~H.}\ \bibnamefont
  {Strogatz}},\ and\ \bibinfo {author} {\bibfnamefont {D.~J.}\ \bibnamefont
  {Watts}},\ }\bibfield  {title} {\bibinfo {title} {Random graphs with
  arbitrary degree distributions and their applications},\ }\href@noop {}
  {\bibfield  {journal} {\bibinfo  {journal} {Physical Review E}\ }\textbf
  {\bibinfo {volume} {64}},\ \bibinfo {pages} {026118} (\bibinfo {year}
  {2001})}\BibitemShut {NoStop}%
\end{thebibliography}%


\begin{thebibliography}{8}%
\makeatletter
\providecommand \@ifxundefined [1]{%
 \@ifx{#1\undefined}
}%
\providecommand \@ifnum [1]{%
 \ifnum #1\expandafter \@firstoftwo
 \else \expandafter \@secondoftwo
 \fi
}%
\providecommand \@ifx [1]{%
 \ifx #1\expandafter \@firstoftwo
 \else \expandafter \@secondoftwo
 \fi
}%
\providecommand \natexlab [1]{#1}%
\providecommand \enquote  [1]{``#1''}%
\providecommand \bibnamefont  [1]{#1}%
\providecommand \bibfnamefont [1]{#1}%
\providecommand \citenamefont [1]{#1}%
\providecommand \href@noop [0]{\@secondoftwo}%
\providecommand \href [0]{\begingroup \@sanitize@url \@href}%
\providecommand \@href[1]{\@@startlink{#1}\@@href}%
\providecommand \@@href[1]{\endgroup#1\@@endlink}%
\providecommand \@sanitize@url [0]{\catcode `\\12\catcode `\$12\catcode
  `\&12\catcode `\#12\catcode `\^12\catcode `\_12\catcode `\%12\relax}%
\providecommand \@@startlink[1]{}%
\providecommand \@@endlink[0]{}%
\providecommand \url  [0]{\begingroup\@sanitize@url \@url }%
\providecommand \@url [1]{\endgroup\@href {#1}{\urlprefix }}%
\providecommand \urlprefix  [0]{URL }%
\providecommand \Eprint [0]{\href }%
\providecommand \doibase [0]{https://doi.org/}%
\providecommand \selectlanguage [0]{\@gobble}%
\providecommand \bibinfo  [0]{\@secondoftwo}%
\providecommand \bibfield  [0]{\@secondoftwo}%
\providecommand \translation [1]{[#1]}%
\providecommand \BibitemOpen [0]{}%
\providecommand \bibitemStop [0]{}%
\providecommand \bibitemNoStop [0]{.\EOS\space}%
\providecommand \EOS [0]{\spacefactor3000\relax}%
\providecommand \BibitemShut  [1]{\csname bibitem#1\endcsname}%
\let\auto@bib@innerbib\@empty
\bibitem [{\citenamefont {Newman}\ \emph {et~al.}(2001)\citenamefont {Newman},
  \citenamefont {Strogatz},\ and\ \citenamefont {Watts}}]{newman2001random}%
  \BibitemOpen
  \bibfield  {author} {\bibinfo {author} {\bibfnamefont {M.~E.}\ \bibnamefont
  {Newman}}, \bibinfo {author} {\bibfnamefont {S.~H.}\ \bibnamefont
  {Strogatz}},\ and\ \bibinfo {author} {\bibfnamefont {D.~J.}\ \bibnamefont
  {Watts}},\ }\bibfield  {title} {\bibinfo {title} {Random graphs with
  arbitrary degree distributions and their applications},\ }\href@noop {}
  {\bibfield  {journal} {\bibinfo  {journal} {Physical Review E}\ }\textbf
  {\bibinfo {volume} {64}},\ \bibinfo {pages} {026118} (\bibinfo {year}
  {2001})}\BibitemShut {NoStop}%
\bibitem [{\citenamefont {Newman}(2018)}]{newman2018networks}%
  \BibitemOpen
  \bibfield  {author} {\bibinfo {author} {\bibfnamefont {M.}~\bibnamefont
  {Newman}},\ }\href@noop {} {\emph {\bibinfo {title} {Networks}}}\ (\bibinfo
  {publisher} {Oxford university press},\ \bibinfo {year} {2018})\BibitemShut
  {NoStop}%
\bibitem [{\citenamefont {Corless}\ \emph {et~al.}(1996)\citenamefont
  {Corless}, \citenamefont {Gonnet}, \citenamefont {Hare}, \citenamefont
  {Jeffrey},\ and\ \citenamefont {Knuth}}]{corless1996lambert}%
  \BibitemOpen
  \bibfield  {author} {\bibinfo {author} {\bibfnamefont {R.~M.}\ \bibnamefont
  {Corless}}, \bibinfo {author} {\bibfnamefont {G.~H.}\ \bibnamefont {Gonnet}},
  \bibinfo {author} {\bibfnamefont {D.~E.}\ \bibnamefont {Hare}}, \bibinfo
  {author} {\bibfnamefont {D.~J.}\ \bibnamefont {Jeffrey}},\ and\ \bibinfo
  {author} {\bibfnamefont {D.~E.}\ \bibnamefont {Knuth}},\ }\bibfield  {title}
  {\bibinfo {title} {On the lambert w function},\ }\href@noop {} {\bibfield
  {journal} {\bibinfo  {journal} {Advances in Computational mathematics}\
  }\textbf {\bibinfo {volume} {5}},\ \bibinfo {pages} {329} (\bibinfo {year}
  {1996})}\BibitemShut {NoStop}%
\bibitem [{\citenamefont {Feigenbaum}(1978)}]{feigenbaum1978quantitative}%
  \BibitemOpen
  \bibfield  {author} {\bibinfo {author} {\bibfnamefont {M.~J.}\ \bibnamefont
  {Feigenbaum}},\ }\bibfield  {title} {\bibinfo {title} {Quantitative
  universality for a class of nonlinear transformations},\ }\href@noop {}
  {\bibfield  {journal} {\bibinfo  {journal} {Journal of statistical physics}\
  }\textbf {\bibinfo {volume} {19}},\ \bibinfo {pages} {25} (\bibinfo {year}
  {1978})}\BibitemShut {NoStop}%
\bibitem [{\citenamefont {Strogatz}(2024)}]{strogatz2024nonlinear}%
  \BibitemOpen
  \bibfield  {author} {\bibinfo {author} {\bibfnamefont {S.~H.}\ \bibnamefont
  {Strogatz}},\ }\href@noop {} {\emph {\bibinfo {title} {Nonlinear dynamics and
  chaos: with applications to physics, biology, chemistry, and engineering}}}\
  (\bibinfo  {publisher} {Chapman and Hall/CRC},\ \bibinfo {year}
  {2024})\BibitemShut {NoStop}%
\bibitem [{\citenamefont {Bonamassa}\ \emph {et~al.}(2025)\citenamefont
  {Bonamassa}, \citenamefont {Gross}, \citenamefont {Kert{\'e}sz},\ and\
  \citenamefont {Havlin}}]{bonamassa2025hybrid}%
  \BibitemOpen
  \bibfield  {author} {\bibinfo {author} {\bibfnamefont {I.}~\bibnamefont
  {Bonamassa}}, \bibinfo {author} {\bibfnamefont {B.}~\bibnamefont {Gross}},
  \bibinfo {author} {\bibfnamefont {J.}~\bibnamefont {Kert{\'e}sz}},\ and\
  \bibinfo {author} {\bibfnamefont {S.}~\bibnamefont {Havlin}},\ }\bibfield
  {title} {\bibinfo {title} {Hybrid universality classes of systemic
  cascades},\ }\href@noop {} {\bibfield  {journal} {\bibinfo  {journal} {Nature
  Communications}\ }\textbf {\bibinfo {volume} {16}},\ \bibinfo {pages} {1415}
  (\bibinfo {year} {2025})}\BibitemShut {NoStop}%
\bibitem [{\citenamefont {Cho}\ and\ \citenamefont
  {Kahng}(2026)}]{cho2026recent}%
  \BibitemOpen
  \bibfield  {author} {\bibinfo {author} {\bibfnamefont {Y.~S.}\ \bibnamefont
  {Cho}}\ and\ \bibinfo {author} {\bibfnamefont {B.}~\bibnamefont {Kahng}},\
  }\bibfield  {title} {\bibinfo {title} {Recent progress on hybrid percolation
  transitions},\ }\href@noop {} {\bibfield  {journal} {\bibinfo  {journal}
  {Entropy}\ }\textbf {\bibinfo {volume} {28}},\ \bibinfo {pages} {68}
  (\bibinfo {year} {2026})}\BibitemShut {NoStop}%
\bibitem [{\citenamefont {Lee}\ \emph {et~al.}(2016)\citenamefont {Lee},
  \citenamefont {Choi}, \citenamefont {Stippinger}, \citenamefont
  {Kert{\'e}sz},\ and\ \citenamefont {Kahng}}]{lee2016hybrid}%
  \BibitemOpen
  \bibfield  {author} {\bibinfo {author} {\bibfnamefont {D.}~\bibnamefont
  {Lee}}, \bibinfo {author} {\bibfnamefont {S.}~\bibnamefont {Choi}}, \bibinfo
  {author} {\bibfnamefont {M.}~\bibnamefont {Stippinger}}, \bibinfo {author}
  {\bibfnamefont {J.}~\bibnamefont {Kert{\'e}sz}},\ and\ \bibinfo {author}
  {\bibfnamefont {B.}~\bibnamefont {Kahng}},\ }\bibfield  {title} {\bibinfo
  {title} {Hybrid phase transition into an absorbing state: Percolation and
  avalanches},\ }\href@noop {} {\bibfield  {journal} {\bibinfo  {journal}
  {Physical Review E}\ }\textbf {\bibinfo {volume} {93}},\ \bibinfo {pages}
  {042109} (\bibinfo {year} {2016})}\BibitemShut {NoStop}%
\end{thebibliography}%

\section*{Author contributions} 
H.J., G.B., and B.M. designed the research and wrote the paper. 
H.J. and B.M. performed analytical and numerical calculations.

\section*{Funding}
This research was supported in part by the National Research Foundation of Korea (NRF) grant funded by the Korea government (MSIT) (No. RS-2025-25433094), and by Global - Learning \& Academic research institution for Master’s · PhD students, and Postdocs (LAMP) Program of the National Research Foundation of Korea (NRF) grant funded by the Ministry of Education (No. RS-2024-00445180).

\section*{Competing interests}
The authors declare no competing interests.

\end{document}


\title{Supplementary Information: Feedback percolation on complex networks}

\author{Hoseung Jang}
\affiliation{Advanced-Basic-Convergence Research Institute, Chungbuk National University, Cheongju, Chungbuk 28644, Korea}
\author{Ginestra Bianconi}
\email{ginestra.bianconi@gmail.com }
\affiliation{Centre for Complex Systems, School of Mathematical Sciences, Queen Mary University of London, Mile End Road, London, E1 4NS, UK}
\author{Byungjoon Min}
\email{bmin@cbnu.ac.kr}
\affiliation{Advanced-Basic-Convergence Research Institute, Chungbuk National University, Cheongju, Chungbuk 28644, Korea}
\affiliation{Department of Physics, Chungbuk National University, Cheongju, Chungbuk 28644, Korea}

\date{\today}
\maketitle

\section{Generating function formulation}

To analyze the size of the giant component in complex networks, we use the generating function framework~\cite{newman2001random,newman2018networks}. 
Let $P(k)$ be the degree distribution of a network. The generating function for this distribution is defined as:
$G_0(x) = \sum_{k=0}^{\infty} P(k) x^k$.
When following an edge to a node, the probability of reaching a node of degree $k$ is proportional to $k P(k)$. The excess degree that is the number of outgoing edges excluding the one we arrived on is generated by $G_1(x) = \frac{G_0'(x)}{G_0'(1)}$.

Let $H_1(x)$ be the generating function for the distribution of the sizes of clusters reached by following a randomly chosen edge. Given an occupation probability $p$ (for either sites or bonds), $H_1(x)$ must satisfy the following self-consistent recursive relation:
\begin{align}
H_1(x) = 1 - p + px G_1(H_1(x)).
\end{align}
Here, $1-p$ accounts for the probability that the edge or the target node is unoccupied, while $px G_1(H_1(x))$ accounts for the case where the connection exists, leading to a node and its subsequent neighbors.

The probability $u$ that a node is not connected to the giant component via a specific neighbor is given by $u = H_1(1)$. 
This leads to the self-consistency equation
\begin{align}
u = 1 - p + p G_1(u).
\end{align}
For bond percolation, $p$ represents the probability that an edge exists. For site percolation, $p$ is the probability that a node is functional.
The fraction of nodes belonging to the giant component, denoted as $S$, is determined by the probability that a randomly chosen node is connected to the giant component through at least one of its neighbors. The fraction of the nodes belonging to the giant component $S$ is:
\begin{align}
    S = 
    \begin{cases} 
    p \left[ 1 - G_0(u) \right] & \text{(Site Percolation)} \\
    1 - G_0(u) & \text{(Bond Percolation)}
    \end{cases}
\end{align}
In site percolation, a node can only belong to the giant component if it is both occupied with probability $p$ and connected to the giant component via its neighbors. In bond percolation, all nodes are present, but edges exist with probability $p$. Therefore, the size $S$ is simply the probability that a node has at least one edge leading to the giant component.

\section{Formal solution of the size of the giant component on ER graphs}

The formal solution of the self-consistency equations can be expressed by 
the Lambert $W$ function \cite{corless1996lambert}, defined as the multivalued inverse of the function 
$f(w) = w\ e^w$, satisfying $W(x)e^{W(x)} = x$.
In our feedback percolation models, the difference equation for $S_n$
can be expressed as:
\begin{align}
S_{n+1} = 1 + \frac{W(-z(p+f(S_n)) e^{-z(p+f(S_n))})}{z(p+f(S_n))}.
\end{align}
This relation allows us to determine the time evolution of the giant component size.

\section{Convergence to the fixed point: linear stability analysis }

Let us consider the system with a feedback, governed by Eqs.~(\ref{Eq_pn}), (\ref{Eq_un}), and (\ref{Eq_sn}). The convergence to the fixed point $u_\infty$ is guaranteed if $|\frac{d u_n}{d u_{n-1}}|_{u_\infty}<1$.
The condition leads to (chain rule)
\begin{align}
\frac{d u_n}{d u_{n-1}} =\frac{d u_n}{d p_{n}} \frac{d p_n}{d u_{n-1}}
\end{align}
Each term in the above equation can be obtained as 
\begin{align}
\frac{d p_n}{d u_{n-1}} &= \frac{d }{d u_{n-1}} f(S_{n-1}) \nonumber\\
&= -f'(1-G_0(u_{n-1}))G'_0(u_{n-1}),\\
\frac{d u_n}{d p_n} &= -1 + G_1(u_n) +p_nG'_1(u_n)\frac{d u_n}{d p_n}.
\end{align}
Defining $\nu=\frac{d u_n}{dp_n}$, we can obtain $\nu$ by solving the following equation:
\begin{align}
\nu = -1 +G_1(u_n)+p_n \nu G'_1(u_n).
\end{align}
Then, $\nu$ is given by
\begin{align}
\nu = -\frac{1-G_1(u_n)}{1-p_n G'_1(u_n)}.
\end{align}
Putting all together, $\left| \frac{d u_n}{d u_{n-1}} \right|$ is finally given by 
\begin{align}
\left|\frac{d u_n}{d u_{n-1}}\right|_{u_\infty} & =\left|-\frac{1-G_1(u_n)}{1-p_nG'_1(u_n)} \times (-f'(1-G_0(u_{n-1}))G'_0(u_{n-1}))\right|_{u_\infty} \nonumber \\
&=\left| \frac{f'(1 - G_0(u_{n-1}))  [1 - G_1(u_n)] G_0'(u_{n-1})}{1 - [p + f(1 - G_0(u_{n-1}))] G_1'(u_n)} \right|_{u_\infty}.
\end{align}

\section{Critical endpoint in positive feedback}
Since $S_\infty=0$ at $p_c=1/z$, the Taylor expansion of $\phi_p(S_\infty,p_c)$ up to second order as follows:
\begin{align}
    \phi_p(S_\infty,p_c) &=1-e^{-zS_\infty(p_c+(1-p_c)S_\infty^{1/q})} \nonumber \\ 
    &\approx zp_c S_\infty + z(1-p_c) S_\infty^{1/q+1} -\frac{1}{2}(zp_c S_\infty + z(1-p_c) S_\infty^{1/q+1})^2 \nonumber\\
    &\approx zp_c S_\infty + z(1-p_c) S_\infty^{1/q+1} -\frac{1}{2}z^2p_c^2 S_\infty^2  \nonumber \\
    &= S_\infty + (z-1)S_\infty^{1/q+1} -\frac{1}{2}S_\infty^2
\end{align}
Then, the second derivative of $\phi_p(S_\infty,p_c)$ at $S_\infty=0$ is given by
\begin{equation}
    \left.\frac{d^2 \phi_p(S_\infty,p_c)}{dS_\infty^2}\right|_{S_\infty=0} = \left.(z-1)\frac{1}{q}\left(\frac{1}{q}+1\right)S_\infty^{1/q-1}\right|_{S_\infty=0} - 1.
\end{equation}
For $q<1$, the first term of the second derivative vanishes, making the second derivative negative. It leads to a stable emergence of a giant component through a continuous transition.
For $q=1$, the second derivative becomes $2z-3$ whose sign depends on the value of $z$.
For $q>1$, the second derivative becomes positive for $z>1$ and negative otherwise. This implies that no small $S_\infty$ solution can exist immediately above $p_c$, thereby resulting in a discontinuous jump for $q>1$.
Therefore, the point $(p,q)=(p_c,1)$ forms a critical endpoint for an ER graph where a non-trivial percolation transition can exist.

\section{Logistic universality class}
We use Feinebaum results \cite{feigenbaum1978quantitative, strogatz2024nonlinear} to prove that for the non-monotonic feedback percolation function,  the observed route to chaos is in the logistic map universality class. Indeed, in this case the effective one dimensional map $S_n=h(S_{n-1})$ describing feedback percolation is unimodal and at its maximum displays a quadratic behavior as it is demonstrated in the following.
The implicit map $S_n=h(S_{n-1})$ of the percolation model with a non-monotonic feedback function is given by:
\begin{align}
S_n = 1 - \exp[- 4 z p S_n S_{n-1}^{1/q}(1-S_{n-1}^{1/q})].
\end{align}
Let $S_n = y$ and $S_{n-1}^{1/q} = x$. The map can be simplified as:
$y = 1 - \exp[- 4 z p y x(1-x)]$.
Taking the first derivative of $y$ with respect to $x$ yields:
\begin{align}
\frac{dy}{dx} = 4zp e^{-4zp yx(1-x)}\left[ \frac{dy}{dx}(x-x^2 )
+ y(1-2x) \right].
\end{align}
Using the relation $dx/dS_{n-1} = \frac{1}{q} S_{n-1}^{(1/q)-1}$, 
the derivative $dS_n/dS_{n-1}$ is obtained as:
\begin{align}
\frac{dS_n}{dS_{n-1}} =  \frac{dx}{dS_{n-1}} \frac{dy}{dx}  = 4zp e^{-4zp yx(1-x)} \left[ \frac{dS_n}{dS_{n-1}}(x-x^2) + y(1-2x) \frac{1}{q} S_{n-1}^{(1/q)-1} \right].
\end{align}
Setting $dS_n/dS_{n-1} = 0$ leads to the critical point $x_c = 1/2$, which corresponds to $S_{n-1} = (1/2)^q$. This confirms that the map has a unique maximum, characterizing it as a unimodal map. 

\begin{figure}
\centering
\includegraphics[width=\textwidth]{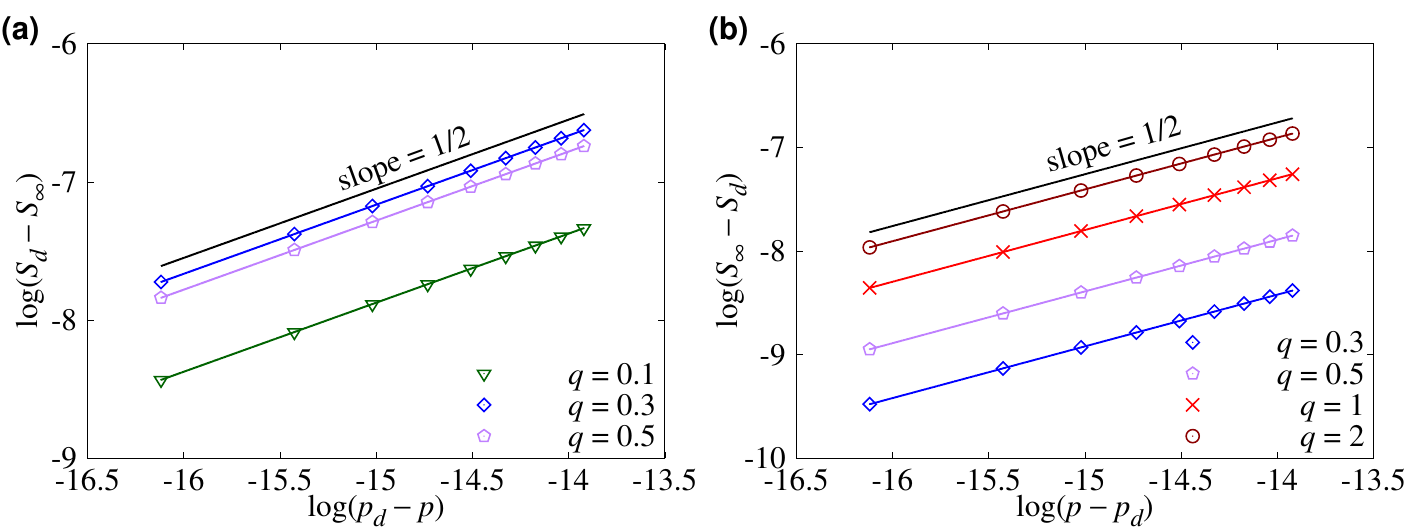}
\caption{\justifying\label{fig_Beta_Bond} 
Scaling behavior of the giant component size.
Log-Log plot showing the scaling behavior of the giant component size near the discontinuous transition point $p_d$ (a) in the positive feedback, and (b) in the size-inverted negative feedback. The black solid line is an eye guide indicating a slope of $1/2$.
}
\end{figure}

In addition, the second derivative of $S_n$ with respect to $S_{n-1}$ can be expressed using the chain rule:
\begin{align}
\frac{d^2 S_n}{dS_{n-1}^2} = \frac{dx}{dS_{n-1}} \frac{d}{dx} \left[ \left( \frac{1}{q} x^{1-q} \right) \frac{dy}{dx} \right].
\end{align}
Since $dy/dx = 0$ at the maximum, the expression simplifies to:
\begin{align}
\frac{d^2 S_n}{dS_{n-1}^2} = \frac{x^{2(1-q)}}{q^2} \frac{d^2 y}{dx^2}.
\end{align}
To evaluate $d^2y/dx^2$, let $A = 4zp e^{-4zp yx(1-x)}$. 
Differentiating the first derivative of $y$ once more, we obtain:
\begin{align}
\frac{d^2 y}{dx^2} = \frac{dA}{dx} \left[ \frac{dy}{dx}(x-x^2) + y(1-2x) \right] + A \left[ \frac{d^2y}{dx^2}(x-x^2) + 2\frac{dy}{dx}(1-2x) - 2y \right].
\end{align}
Applying the conditions for the critical point, 
$x_c = 1/2$ and $dy/dx|_{x=x_c} = 0$, the equation reduces to:
\begin{align}
\frac{d^2 y}{dx^2} = A \left[ \frac{1}{4} \frac{d^2y}{dx^2} - 2y \right].
\end{align}
Solving for $d^2y/dx^2$ at the maximum yields:
\begin{align}
\left. \frac{d^2 y}{dx^2} \right|_{x=x_c} = \frac{-2Ay}{1-A/4},
\end{align}
which is non-zero. 
This non-vanishing second derivative confirms that the map possesses a quadratic 
maximum, leading to the transition to chaos in this system within 
the logistic universality class.

\section{Hybrid phase transitions}

\begin{figure}
\centering
\includegraphics[width=0.6\textwidth]{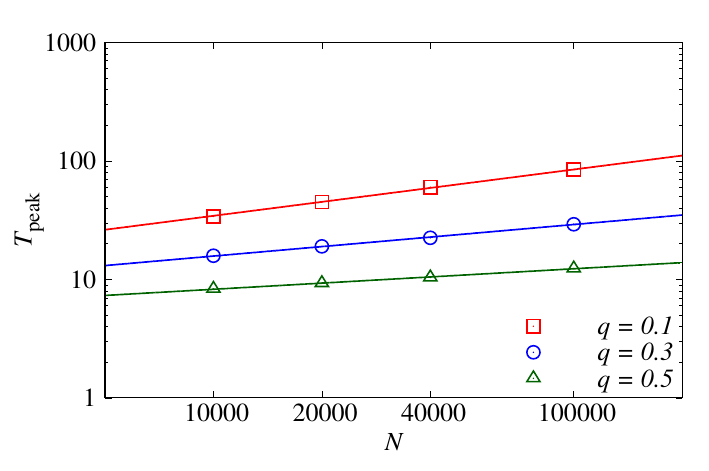}
\caption{\justifying\label{fig_NOI_Peak} 
Divergence of the number of iterations peak height.
Log-Log plot showing the divergent behavior of the number of iterations (NOI) peak height for feedback strength $q=0.1, 0.3$, and $0.5$ in the positive feedback.
}
\end{figure}

In this section, we show that the discontinuous jump in $S_\infty$ at $p_d$ has a hybrid nature \cite{bonamassa2025hybrid,cho2026recent,lee2016hybrid}.
Let us start with the self-consistency equation at a stable fixed point $S_\infty$ of the size-inverted negative feedback model:
\begin{equation}
    S_\infty=1-e^{-zpS_\infty^{1/q+1}}.
\end{equation}
Then, we define $F(S_\infty,p)$ as following:
\begin{equation}
    F(S_\infty,p) \equiv 1-e^{-zpS_\infty^{1/q+1}}-S_\infty =0.
\end{equation}
By performing a Taylor expansion of $F(S_\infty,p)$ up to second order near the transition point $(S_d,p_d)$, where
\begin{equation}
    F(S_d,p_d)=0, \quad \left.\frac{\partial F}{\partial S_\infty}\right|_{S_d,p_d}=0
\end{equation}
and removing vanishing terms, we obtain the following expression for $F(S_\infty,p)$:
\begin{align}
    F(S_\infty,p) &\approx \left.\frac{\partial F}{\partial p}\right|_{S_d,p_d}(p-p_d) +\frac{1}{2}\left.\frac{\partial^2 F}{\partial S_\infty^2}\right|_{S_d,p_d}(S_\infty-S_d)^2 \nonumber \\
    &+ \left.\frac{\partial^2 F}{\partial S_\infty \partial p}\right|_{S_d,p_d}(S_\infty-S_d)(p-p_d) =0.
\end{align}
Since $(S_\infty-S_d)\ll 1$, the third term becomes negligible compared to the first term. Therefore,
\begin{equation}
     S_\infty - S_d\sim(p-p_d)^{1/2},
\end{equation}
which indicates a hybrid phase transition with the critical exponent $\beta =1/2$. Similarly, the scaling behavior $S_d-S_\infty \sim (p_d-p)^{1/2}$ is observed in the positive feedback model. Figure~\ref{fig_Beta_Bond} shows the scaling behaviors in the positive and size-inverted negative feedback models.

We also show that the number of iterations (NOI) in the positive feedback exhibits a pronounced peak at the discontinuous transition point. 
Here, we examine the dependence of the peak height on the system size $N$. 
As shown in Fig.~\ref{fig_NOI_Peak}, the peak height increases systematically with the size $N$, 
indicating that it diverges in the thermodynamic limit ($N \rightarrow \infty$).
This behavior confirms that the observed peak is not a finite-size artifact but is associated with the discontinuous phase transition.

\begin{figure}
\centering
\includegraphics[width=1.0\linewidth]{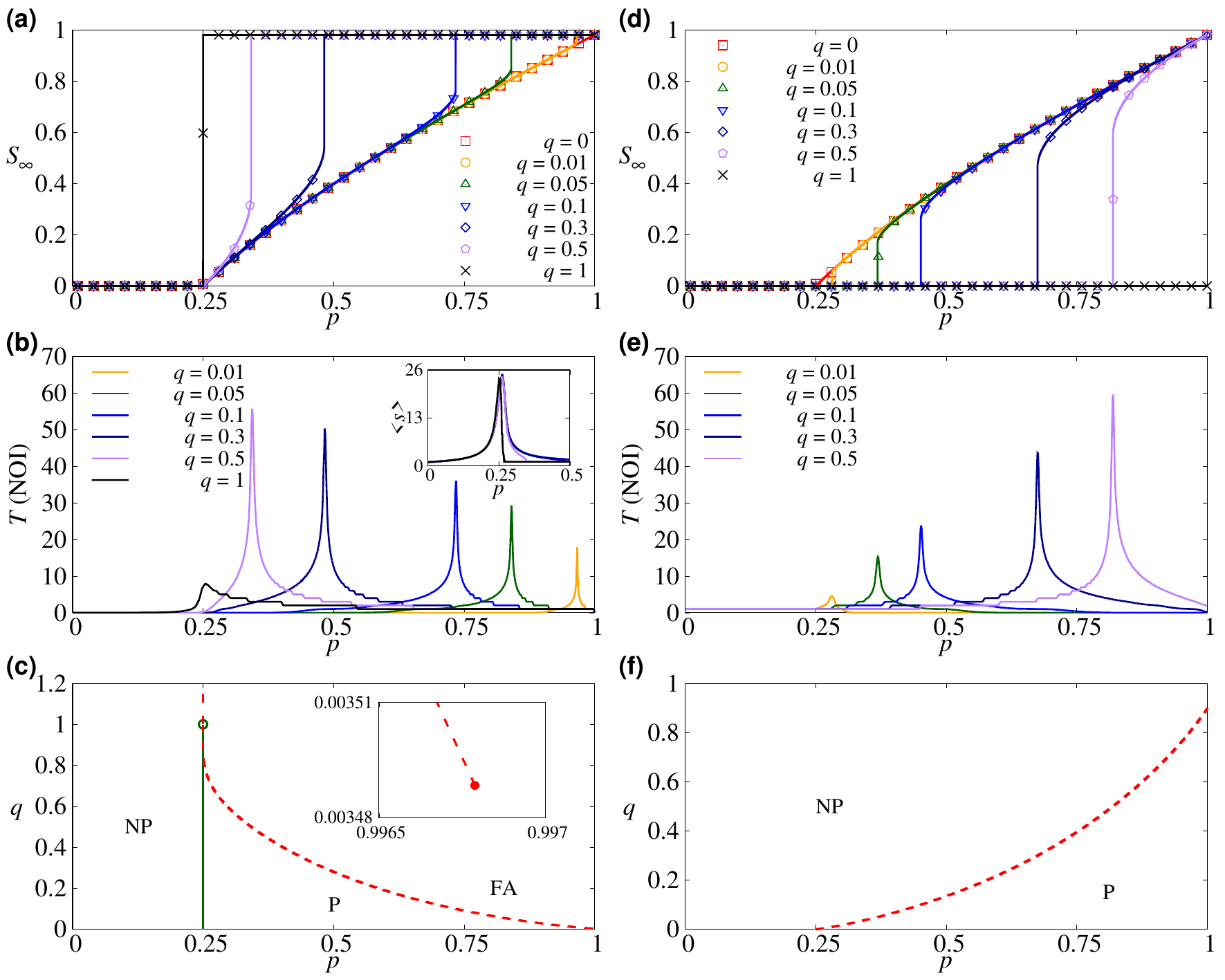}
\caption{\justifying\label{fig_Both_Site}
Phase transitions in site percolation with feedback.
(a, d) The size of the giant component $S_\infty$ of the site percolation under the positive and size-inverted negative feedback as a function of probability $p$ on an Erd\H{o}s-R\'enyi (ER) graph with average degree $z=4$. Symbols represent numerical results for system size $N=10^5$, and solid lines represent theoretical predictions. (b, e) The number $T$ of iterations (NOI) required to reach the steady state in numerical simulations as a function of $p$. The inset in panel (b) displays the average size of finite connected components. (c, f) Phase diagram in the $(p,q)$ plane. The solid line indicates the boundary of continuous transitions, and the dashed line indicates that of a discontinuous jump. The inset in panel (c) shows the critical point $(p_{cp},q_{cp})$ for the positive feedback model.
Here, $\mathbf{NP}$, $\mathbf{P}$, and $\mathbf{FA}$ denote the non-percolating, percolating, and fully activated phases, respectively.
}
\end{figure}

\section{Site percolation with feedback}

We also analyze site percolation with feedback, where the activation probability of sites depends dynamically on the size of the giant component.
In this section, we examine site percolation in the presence of positive feedback, $f(S_n)=S_n^{1/q}$. For an ER graph with mean degree $z$, the self-consistency equation for the steady state where $S_{n-1}=S_n=S_\infty$ is given by $S_\infty = (p + S_\infty^{1/q})(1 - e^{-zS_\infty})$. Due to the physical constraint that $p + S_{\infty}^{1/q}$ must be in the range $[0, 1]$, we set $p + S_{\infty}^{1/q} = 1$ whenever it exceeds unity. As shown in Fig.~\ref{fig_Both_Site}(a), the system exhibits diverse phase transition behaviors, similar to those observed in bond percolation. The network shows a continuous phase transition at $p_c = 1/z$, which may be followed by a subsequent discontinuous transition at $p_d>p_c$. Here, the two transitions at $p_c$ and $p_d$ are also driven by distinct mechanisms. The average size $\langle s \rangle$ of finite components and the NOI $T$ show a divergence at $p_c$ and $p_d$, respectively, as shown in Fig.~\ref{fig_Both_Site}(b). The phase diagram is depicted in Fig.~\ref{fig_Both_Site}(c). The percolation transition between NP and P occurs at $p_c=1/z$ for any value of $q$. However, the transition is continuous for $q<1$ and discontinuous for $q\ge1$. A discontinuous jump in the size of the giant component occurs at a higher value of $p_d$ as $q$ decreases, and it vanishes at the critical point $(p_{cp},q_{cp}) \approx (0.9968, 0.0035)$.

We now turn to the size-inverted negative feedback model $f(S_n)=-(1-S_n)^{1/q}$, and examine how this feedback affects the percolation process. Following the same analytical framework as before, we calculate the size of the giant component and the NOI, varying the feedback strength $q$. The results are shown in Fig.~\ref{fig_Both_Site}(d) and (e). In the size-inverted negative feedback model, the system exhibits only a discontinuous phase transition, and the NOI shows a clear peak at the transition point, as in the positive feedback case.
Contrary to the positive feedback case, the transition point shifts to the right and then disappears as the feedback strength increases. This correlation between the feedback strength and the transition point is also clearly shown in the phase diagram (Fig.~\ref{fig_Both_Site}(f)). 

\section*{Supplementary References}

\bibliography{feedback}